# A novel approach to assess hydrogen embrittlement (HE) susceptibility and mechanisms in high strength martensitic steels

Tuhin Das[1†], Salim V. Brahimi[1,2], Jun Song[1], and Stephen Yue[1†]

*1 Department of Materials Engineering, McGill University, Montreal, Quebec H3A 0C5, Canada*

*2 Industrial Fasteners Institute, Cleveland, OH 44131, USA*

## Abstract

A rapid fracture test in four-point bending is proposed to assess hydrogen embrittlement (HE) susceptibility of high strength martensitic steels. The novelty of this technique is the rapid rate of loading, whereas conventional approaches require prolonged slow strain rate testing. The essential fractographic features required to identify the mechanisms of HE failure remain evident, despite the fast loading conditions. To demonstrate these attributes, two quenched and tempered steels at two different strength levels were tested, with and without pre-charging of hydrogen. Stress coupled hydrogen diffusion finite element analysis was performed to calculate both stress and hydrogen concentration distributions. In addition to fractographic analysis, a mechanistic description rooted in hydrogen enhanced decohesion (HEDE) mechanism was used to corroborate the mechanical test data. The study shows that the approach is capable of quantifying HE susceptibility by being responsive to key factors affecting hydrogen induced fracture, thus developing further understanding on the HE of martensitic steels.

[†] *Corresponding* Authors
*Email addresses*: tuhin.das@mail.mcgill.ca (T. Das), steve.yue@mcgill.ca (S. Yue).

**Keywords:** Hydrogen embrittlement, fast fracture, finite element analysis, HE mechanism, material susceptibility, martensitic steels

## 1. Introduction

Quench and tempered (Q & T) martensitic steels are one of the most important structural materials used today. The applications of these steels involve a wide range of industries starting from fastener, oil & gas to automotive, aerospace and nuclear [1–5]. The ingress of hydrogen into these metallic systems from service environments, or during manufacturing, can result in loss of integrity and catastrophic failures threatening their reliability and durability. For instance, high strength Q & T steels, known for their crucial applications in aerospace industries such as landing gears, and structural applications involving high strength fastener components, can suffer from premature failures due to hydrogen embrittlement (HE). The HE cracking of high strength anchor rods and structural bolts in the San Francisco–Oakland Bay Bridge is a notable incident [6,7]. However, martensitic steels are not yet well understood to date because of their complex microstructure [8]. Thus, understanding the hydrogen (H) induced fracture mechanism(s) of these materials is challenging in spite of the availability of extensive literature on their susceptibility [1,9–14]. In general, the susceptibility of a Q & T steel to HE could be related to the mechanical properties such as yield strength, toughness, hardness etc. A decrease in resistance to HE failures with the increase in hardness and yield strength is well accepted in the case of high strength steels [1,15–17]. The role of strength in affecting the stress gradient, the driving force for H diffusion has been explored widely as opposed to its connection to the mechanics of fracture in the presence of H. In general, fracture toughness is known to decrease with the increase in strength, but in the presence of H, the fracture toughness measurements strike a lowest point when a critical yield strength level is exceeded [18]. However, the reason for such behavior was not clearly elucidated. The influence of a certain hydrogen content on different steel grades of the same strength level also needs further understanding. Again, it is well known that microstructures mandate the H diffusion and trapping characteristics inside the materials [9,19–21], in addition to its contribution to strength. As a matter of fact, H diffusivity being influenced by microstructural features resulting in an apparent diffusivity can be pointed as a "global" role of the microstructure. However, microstructural participation in the micromechanics of HE failure and ultimately affecting material susceptibility is not very well understood and needs further attention. In this study, a combined approach based on experimental investigations and finite element analysis (FEA) has been adopted. The experimental investigation is based on a newly proposed fast fracture test methodology that generates percent notch fracture strength (NFS%) value. The conventional methodologies such as ASTM F1624 standard test, linearly increasing stress test (LIST), constant extension rate test (CERT) etc. are already available to evaluate HE susceptibility. [22,2,23]. These test methods are also advantageous in a way that they are easy to set up and does not involve the complexity and cost related to a fracture mechanics based approach. However, there are some limitations and critical issues with the above methodologies. Some of the limitations of the slow strain rate tensile testing (SSRT) methods, involving their HE susceptibility and subcritical cracking prediction capabilities are well reported in [24]. On the other hand, the different hold times at a particular load level, prescribed in ASTM F1624 in response to different hardness levels could be a matter of concern. Additionally, the time to evaluate the HE susceptibility metrices is also quite significant, using each of the above test methods. For instance, utilizing some of the available and developed incremental step load (ISL) test protocols based on ASTM F1624 are notably slow, consuming 15 to 48 hours for each test [1,13,25]. In contrast, the proposed test method involves cathodic pre-charging of H followed by a four-point bend test at a relatively higher loading rate of 445 N/min, which remarkably reduces the test time. But, the main advantage of this innovative approach includes the straightforward evaluation of HE susceptibility based on similar exposure levels to hydrogen in all cases, owing to the precharging of hydrogen. Another novelty is that it takes out the time factor involved with long range H diffusion to the stress concentrated region during the evaluation process. This particular feature facilitates to exclusively study the embrittlement phenomenon from a "local" perspective i.e. the interaction of H with the local stress-strain fields, provided that sufficient amount of H distribution takes place in the notch tip area through precharging. Therefore, the results necessary reflects the intrinsic HE susceptibility of a particular material, without entangling much influence of global microstructure on H diffusion and fracture. Hence, the present study discusses the potential of establishing a new test technique to determine HE susceptibility of steels in a simple and much faster way. Nevertheless, it also depends on the priority of an examiner to include or preclude certain factors while evaluating HE susceptibility of a material. Although, the current approach can be typically sorted as an internal hydrogen embrittlement evaluation method, the values and trends of the results are roughly comparable to both environmental and internal HE test methods [1,16]. However, an exhaustive comparison is not the aim of the current study; a subsequent study will be carried out in this direction. The applicability of the proposed test method to other material systems would also need further examination. Comprehensive fracture surface mapping is also carried out in this study to discuss the mechanisms of HE failure being generated by the mechanical test technique in each of the Q & T martensitic steels. At the same time, the numerical study contributes towards the estimation of necessary stress and H concentration distribution profiles related to HE failure.

Finally, a mechanistic description rooted in hydrogen enhanced decohesion (HEDE) mechanism, and dislocation emission phenomenon from a crack tip, based on the formulation of Gerberich and coworkers [26–28], has been proposed to provide further insights into the HE failure of Q & T martensitic steels. The outcomes from the present study can potentially provide necessary guidelines towards the development of new HE resistant high strength martensitic steels.

## 2. Methodology

*2.1. Materials*

Two different alloys AISI 4340 and 4140 were selected for this study. Table 1 gives the compositions of the steel grades. All the materials, 4340, 4340L and 4140 were manufactured, heat treated and supplied by the industries: TATA steel (UK) & Certified Metal Craft Inc. (US), Cetim Materials (France) and US Bolt Manufacturing Inc. & American Petroleum Institute (US) respectively. Table 1 also lists the average hardness values of the materials in Rockwell C scale (HRC). The tensile properties of the materials were obtained from uniaxial tensile tests following the ASTM standard E-8 [29]. Please note that "L" in 4340L represents lower hardness or strength of the heat as compared to the other heat 4340.

Machining was performed as per ASTM F519 type 1e specifications to obtain the specimens in the form of square single edge notched bars [25,30]. The notch was created by wire electrical discharge machining within 0.5mm of the final notch depth while the surface finish of the notch was obtained in the order of 0.4 μm root square mean. The specimen dimensions and final notch depth were obtained by a low stress grinding process. After the final machining, no cleaning (mechanical or chemical) was performed, only a stress relief heat treatment was performed.

Hardness was measured using a Rockwell hardness tester with a diamond tip (C-Scale), from four different locations in the mid-section area of the square notch bars.

**Table 1**

Chemical composition of the Steel grades (wt.%).

| Steel Grade | C | Mn | Ni | Cr | Mo | Si | S | P | Cu | Al | Hardness (HRC) |
|---|---|---|---|---|---|---|---|---|---|---|---|
| **4340** | 0.40 | 0.75 | 1.78 | 0.83 | 0.25 | 0.25 | 0.001 | 0.005 | 0.12 | 0.03 | 52.2 |
| **4340L** | 0.42 | 0.75 | 1.73 | 0.80 | 0.23 | 0.22 | 0.006 | 0.009 | 0.23 | - | 42.8 |
| **4140** | 0.41 | 0.97 | - | 1.03 | 0.20 | 0.23 | 0.003 | 0.011 | - | - | 41.7 |

*2.2. Hydrogen charging and fast fracture testing*

A fast loading rate as compared to the conventional test methods was selected to perform the susceptibility tests. The test equipment used for this investigation was an RSL® loading frame [31]. A computer-controlled four-point bend load control frame that can reach the target loads with an accuracy of ± 1.78 N (0.4 lb) and can hold displacement within ± 0.13 µm was used. A fast fracture test methodology without any precharging of H was considered to determine the baseline load measurements. A continuous loading rate of 445 N/min (100 lbs/min) was applied until the specimen fractured and the load at fracture was noted. This process was repeated at least three times and an average value of the fast fracture load was reported for each grade. The same loading rate was used to test the steel samples precharged with H and the percentage notch fracture strength (NFS$_%$) was evaluated as the measure of susceptibility. The samples were pre-charged with H using a 5% (volume/volume) sulphuric acid solution containing 1 g/L of thiourea, which was added as a poison to avoid the formation of molecular hydrogen through the recombination of H atoms [32]. The hydrogen charging process was carried out under galvanostatic control by applying a constant current of 20 mA, i.e. approximately 0.821 mA/cm$^2$, current density with the help of a potentiostat using an electrolytic cell. It should be noted that the choice of current density was maintained at a lower value as mentioned in [32], to avoid any damage to the samples because of the severity of charging process. The steel specimen was used as the working electrode, while a saturated calomel electrode (SCE) and a platinum wire of length 10 cm and 1 mm diameter were used as the reference and counter electrode respectively. It should be noted that the samples were immediately tested within a time span of 2-3 mins to minimize any H loss during the intermediate transfer process. Finally, the end of the test was identified by a load drop indicating the onset of cracking, and the NFS$_\%$ is given by Eq. (1) as:

$$Percent\ notch\ fracture\ strength\ (NFS_\%) = \frac{Fast\ fracture\ strength\ in\ presence\ of\ H}{Fast\ fracture\ strength\ in\ air} \times 100 \quad (1)$$

At least three tests were performed to obtain an average percent NFS (NFS$_\%$) value and the fracture surfaces were preserved for fractographic analysis using FEI Quanta 450 Environmental Scanning Electron Microscope and Hitachi SU-3500 Variable Pressure-SEM, at high vacuum.

*2. 3. Finite element Analysis (FEA): Hydrogen diffusion under the influence of stress*

The FEA model mainly comprises of two physical models, a diffusion model coupled with a stress analysis model. In this section, a brief overview on the modeling approach has been presented. At a constant temperature, applying the mass conservation law, the rate of change of lattice H concentration inside an arbitrary domain of volume $\Omega$ bounded by a surface $\partial\Omega$ can be written as:

$$\frac{\partial}{\partial t}\int_\Omega C_L\ d\Omega + \int_{\partial\Omega} \boldsymbol{\varphi}.\boldsymbol{n}\ d\partial\Omega = 0 \quad (2)$$

where $C_L$ is the lattice hydrogen concentration per unit volume. $\varphi$ denotes hydrogen flux vector and **n** is an outward-pointing unit normal vector. Assuming, there is no interaction between the diffusion species due to low concentration, the rate of diffusion entirely depends upon the H mobility [33]. Therefore, the hydrogen flux $\varphi$ can be written as:

$$\varphi = -\frac{D_L C_L}{RT}\nabla\mu \tag{3}$$

where $D_L$ is the lattice diffusivity, $\mu$ is the chemical potential, $R$ is the Gas constant (8.314 J.mol$^{-1}$K$^{-1}$) and $T$ is the temperature (in Kelvin). At constant temperature and pressure, the chemical potential for a system under external stress and could be written as [34]:

$$\mu = \mu_0 + RT\ln\frac{C_L}{N_L} - \sigma_h V_H \tag{4}$$

where $\mu_0$ is the chemical potential at a reference state, $N_L$ is the number of lattice (or interstitial) sites per unit volume [35], $\sigma_h$ is the hydrostatic stress (i.e. $\sigma_h = \frac{1}{3}\sum_1^3 \sigma_{nn}$, $\sigma_{nn}$ = normal stress components) and $V_H$ is the partial molar volume of hydrogen. Substituting Eq. (4) in Eq. (3), we get the expression of flux as:

$$\varphi = -D_L\nabla C_L + \frac{D_L C_L V_H}{RT}\nabla\sigma_h \tag{5}$$

Further, by substituting the expression of $\varphi$ from Eq. (5) in Eq. (2) and applying the divergence theorem considering a continuous arbitrary volume $\Omega$, the lattice H diffusion equation can be obtained as:

$$\frac{\partial C_L}{\partial t} - \nabla \cdot (D_L \nabla C_L) + \nabla \cdot \left(\frac{D_L C_L V_H}{RT}\nabla\sigma_h\right) = 0 \tag{6}$$

A weak formulation of the H diffusion problem along-with the complete formulation of the stress coupled transport-mechanical scheme has been documented in a preceding study [25]. Again, hydrogen is known to get trapped due to plastic strain [35]. Now, assuming Oriani's theory of local equilibrium [36] between H in lattice sites and trap sites of a particular type such as dislocations in this case, we can write:

$$\frac{(1-\theta_L)\theta_T}{(1-\theta_T)\theta_L} = K_{eq} = e^{\frac{\Delta G_T}{RT}} \qquad (7)$$

where $K_{eq}$ is termed as the trap equilibrium constant, $\Delta G_T$ is the trap binding energy, $\theta_L$ $(= C_L/N_L)$ and $\theta_T$ $(= C_T/N_T)$ are the lattice and trap site occupancies respectively. Due to low solubility of H in martensitic steels (i.e. $\theta_L << 1$) and using Eq. (7), the trapped H concentration, $C_T$ can be written as:

$$C_T = \left(\frac{K_{eq}N_T C_L}{\{N_L + K_{eq}C_L\}}\right) \qquad (8)$$

where $N_T$ is the trap site density per unit volume. The change in dislocation density, $\rho$, measured in dislocation line length per unit volume with plastic strain, $\varepsilon_p$ could be related using Eq. (9) [37,38] as:

$$\rho = \begin{cases} \rho_0 + 2\tau\varepsilon_p, & \varepsilon_p \leq 0.5 \\ \rho_0, & \varepsilon_p > 0.5 \end{cases} \qquad (9)$$

where $\rho_0 = 10^{10}$ line length/m² is the dislocation density in an annealed material and $\tau = 10^{16}$ line length/m². Again, the dislocation trap density is related to dislocation density by Eq. (10):

$$N_T = \sqrt{2}\,\rho/a_0 \qquad (10)$$

where $a_0$ is the lattice parameter of bcc iron with a value of 0.287 nm [39,40]. Therefore, by substituting $N_T$ from Eq. (10) in Eq. (8), the *equilibrium* H concentration trapped due to plastic strain can be calculated. In this study, all the calculations were performed assuming that a site (lattice or trap) could be occupied by only one H atom.

The elastic-plastic behavior according to J2 plasticity has been considered as material response [41]. A 2D plane strain condition with an isotropic strain hardening process has been assumed to carry out the simulations. Therefore, the weak formulation accounting the elastoplastic deformation of the material could be obtained by applying the principle of virtual work [42]. In this study, the effect of hydrogen on macroscopic stress-strain relation of the materials has been neglected, based on the fact that no significant influence of H on the constitutive behavior of Q & T steels was observed [43–45].

*2. 3.1 Boundary/initial conditions and implementation of the model*

A homogeneous distribution of hydrogen concentration, $C_0$ throughout the domain was considered as the initial condition for each simulation. A zero-flux condition was considered at the sample boundaries to simulate the experimental condition. The equivalency of gaseous and electrolytic charging condition has been earlier summarized in by Liu and Atrens [46]. Now, the equilibrium H concentration during the galvanostatic charging process could be approximated using the relation between the constant current density, $j_\infty$ (i.e. 0.821 mA/cm$^2$) and H concentration, $C_0$ given by Eq. (11) [46,47] as:

$$C_0 = \frac{j_\infty l}{nFD_L} \qquad (11)$$

where $l$ is the specimen thickness (i.e. the notch length of 6.4 mm), n is the number of electrons transferred (i.e. 1 equivalent), F is the Faraday constant (96,500 C/mol/equivalent). Substituting the value of lattice H diffusivity, $D_L$ = 1.5E-08 m$^2$/s [48] in Eq. (11), 36.305 mol/m$^3$ (~4.6 wt. ppm) of hydrogen concentration was evaluated. The estimated concentration will be considered as the initial condition for the FEA simulations. This estimation is based on some simplified assumptions that will be helpful in capturing trends while performing the calculations, and is also in good agreement with the value for gaseous hydrogen charging in [49].

The implementation of the FEA model in the commercial program ABAQUS, has been carried out following the analogy comparing heat transfer and mass diffusion as documented in the papers [50,51]. In order to implement the stress coupled diffusion equation (Eq. (6)) in ABAQUS, writing user subroutine is necessary. A user subroutine UMATHT was thus developed and the analogy among the variables in heat transfer and diffusion based on Table 2 was used. The spatial gradient of lattice H concentration i.e. $\nabla C_L$ is available in basic UMATHT interface. However, the gradient of hydrostatic stress i.e. $\nabla \sigma_h$ also needs to be supplied into the UMATHT subroutine for the calculation of diffusion flux as per Eq. (5). The calculation of $\nabla \sigma_h$ was carried out using a finite element code developed in the computational programming environment, MATLAB. The finite element procedure involving the evaluation of $\nabla \sigma_h$ has been briefly discussed here. First, the hydrostatic stresses at the nodal points were extracted following the stress analysis performed in ABAQUS from the output file using a Python script code. After the extraction process, the gradient of hydrostatic stress at a point, $[\nabla \sigma_h]_{x,y}$ in 2D can be calculated by using the idea of gradient matrix $[G]$ such that:

$$[\nabla \sigma_h]_{x,y} = [G] [\sigma_h]_e \tag{12}$$

where, $[\sigma_h]_e$ is the vector containing the hydrostatic stress at the nodal points in an element. Thus, considering four nodal points and shape functions, $N_i$, Eq. (12) for a quadrilateral element can be rewritten as:

$$\begin{bmatrix} \frac{\partial \sigma_h}{\partial x} \\ \frac{\partial \sigma_h}{\partial y} \end{bmatrix}_{x,y} = \begin{bmatrix} \sum_{i=1}^{4} \frac{\partial N_i}{\partial x} \sigma_{h,i} \\ \sum_{i=1}^{4} \frac{\partial N_i}{\partial y} \sigma_{h,i} \end{bmatrix} \tag{13}$$

where the derivatives of the shape functions with respect to the global coordinates (x, y) could be obtained by using the Jacobian matrix. Using chain rule, we can convert the derivatives of the shape functions with respect to the local coordinates (ξ, η) in terms of global coordinates as:

$$\begin{bmatrix} \frac{\partial N_i}{\partial \xi} \\ \frac{\partial N_i}{\partial \eta} \end{bmatrix} = \begin{bmatrix} \frac{\partial x}{\partial \xi} & \frac{\partial y}{\partial \xi} \\ \frac{\partial x}{\partial \eta} & \frac{\partial y}{\partial \eta} \end{bmatrix} \begin{bmatrix} \frac{\partial N_i}{\partial x} \\ \frac{\partial N_i}{\partial y} \end{bmatrix} = [J] \begin{bmatrix} \frac{\partial N_i}{\partial x} \\ \frac{\partial N_i}{\partial y} \end{bmatrix} \quad (14)$$

Again, for an isoparametric system, the Jacobian matrix *[J]* can be written as:

$$[J] = \begin{bmatrix} \sum_{i=1}^{4} \frac{\partial N_i}{\partial \xi} x_i & \sum_{i=1}^{4} \frac{\partial N_i}{\partial \xi} y_i \\ \sum_{i=1}^{4} \frac{\partial N_i}{\partial \eta} x_i & \sum_{i=1}^{4} \frac{\partial N_i}{\partial \eta} y_i \end{bmatrix} \quad (15)$$

Therefore, by calculating the Jacobian matrix using Eq. (15) and using the inverse of it in Eq. (14), we can evaluate the gradient matrix. Finally, utilizing the gradient matrix, the gradient of hydrostatic stress is calculated from Eq. (13) for an element. Similarly, by running a loop over of all the elements, the distribution of the gradient of the hydrostatic stress over a geometry can be obtained. These values were again utilized in the UMATHT subroutine for H concentration calculations. The main code was written in MATLAB, constituting all the other auxiliary programs to be called and executed in a sequence from MATLAB.

**Table 2**

Analogy among the variables in heat transfer and diffusion.

| Heat transfer | | Diffusion | |
|---|---|---|---|
| Variable | *Symbol* | Variable | *Symbol* |
| Temperature | $\theta$ | Lattice H Concentration | $C_L$ |
| Thermal Energy | $U$ | Total H concentration | $C$ |

| Density | $\rho$ | 1 | -- |
| Conductivity | $k$ | Lattice H diffusivity | $D_L$ |
| Specific heat | $\frac{\partial U}{\partial \theta}$ | 1 | $\frac{\partial C_L}{\partial C}$ |
| Thermal flux | $-k\nabla\theta$ | Diffusion flux | $\phi$ |

A mesh study was performed to obtain the optimum number of elements. The geometry was then meshed with a total number of 8986 nodes and 8707 elements, with 100 elements being uniformly distributed at the notch tip. The finite element model domain, which is a single-edge notched SE(B) specimen is schematically illustrated in Fig. 1. The specimen dimensions are identical to the experimental test samples i.e. ASTM F519 type 1e specifications. The length (L) of the specimen is 55.9mm, while both the thickness (B) and width (W) are the same i.e. 10mm, with a notch tip radius (r) of 0.25 mm. In Fig. 1, $\ell$ and $\hbar$ represent the concentrated load points and pinned points respectively and a loading rate ($\dot{P}$) of 81.5 N/s, calculated from the fast fracture test was used to perform the finite element calculations. The plane strain quadrilateral elements (CPE4) and the quadrilateral heat transfer elements (DC2D4) available in ABAQUS were used for stress and diffusion analyses respectively. The 2D diffusion FEA model was first utilized to calculate the time at which the sample geometry reaches equilibrium with the surface concentration, 36.305 mol/m$^3$. The result (Appendix A) shows that achieving a uniform distribution in the notch root area with the surface concentration in 1 hour of charging is well feasible. It further justifies the assumption to consider an initial condition, C$_0$ to perform the stress coupled simulations. Since, relatively fast loading rates were considered, *"quasi-equilibrium"* H concentrations (concentration distribution immediately after 1s), or simply referred to be as lattice H concentrations were evaluated and reported. In addition to quasi-equilibrium H concentration, *"equilibrium"* H concentrations were also evaluated for validation purposes. The equilibrium H concentration further closely resembles the H distribution during conventional slow strain rate testing. A total time for equilibrium lattice H concentration calculations i.e. 900s was employed based on the guidelines given in [51] with a maximum time step of 50s. The SI (mm) unit system was considered for the calculations in ABAQUS and the temperature for all the calculations were maintained at 293 K. It is also necessary to specify the value of the maximum change in concentration at each increment, indicated by a variable called *deltmx* in ABAQUS. The minimum integer value of *deltmx* = 1 molmm$^{-3}$ has been used here.

All the necessary materials parameters are listed in Table 3, and the stress-strain curves from the tensile tests were modeled using the following strain hardening functions are given below. The constitutive behavior for 4340 was adopted from [48] following the similarity in hardness and other properties.

4140:

$$\sigma_0(1 + 3.13\varepsilon_p + 7.17\varepsilon_p^2)$$

4340L:

$$\sum_{i=1}^{3} \left(\frac{U_i}{V_i}\right) * \{1 - e^{(-V_i * \varepsilon_p)}\}$$

4340 [48]:

$$\sigma_0 \left(1 + Y \frac{\varepsilon_p}{\sigma_0}\right)^{1/n}$$

where $\sigma_0$ is the yield strength, $\varepsilon_p$ is the plastic strain, $n$ = 7.7 is the strain hardening exponent, and $U_1$ = 800MPa, $U_2$ = 1792 MPa, $U_3$ = 60000 MPa, $V_i$ are the constants with values 4, 80, 1000 respectively.

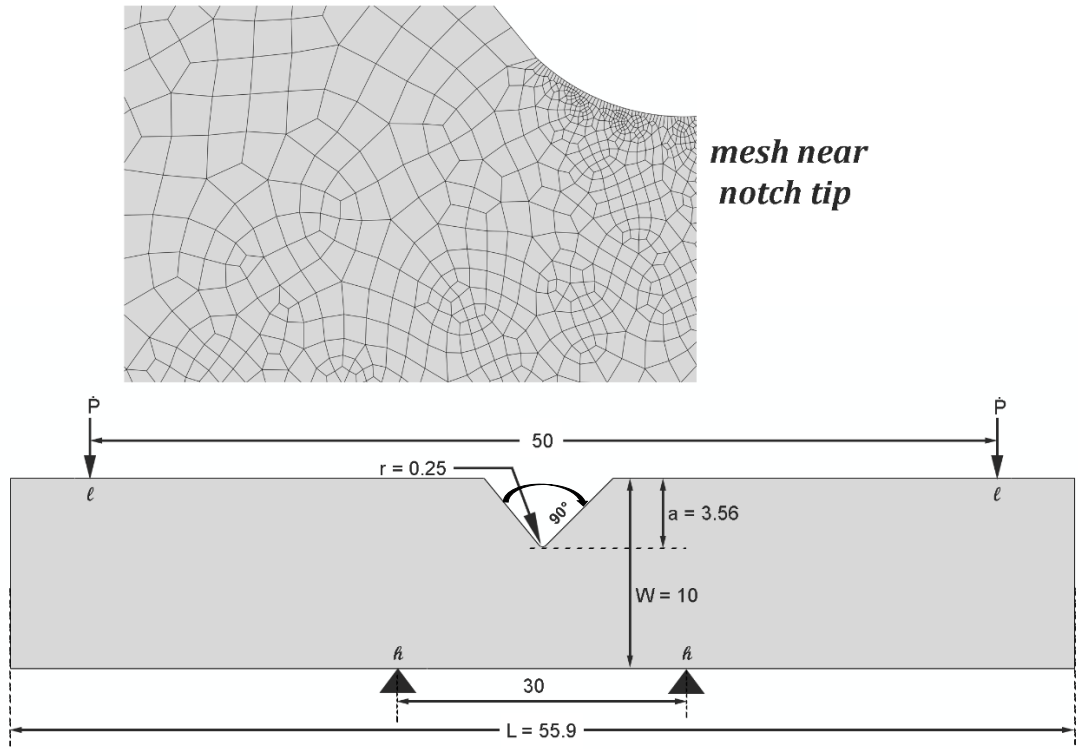

**Fig. 1.** Schematic of the sample geometry and the finite element model domain. All dimensions are in mm.

**Table 3**

Material Parameters

| Parameters | Symbols | 4340 | 4340L | 4140 |
|---|---|---|---|---|
| Elastic modulus (GPa) | $Y$ | 200 | 211 | 211 |
| Poisson ratio | $v$ | 0.3 | | |
| Density (kg/m³) | $\rho$ | 7850 | | |
| Yield strength (MPa) | $\sigma_0$ | 1500 | 1270 | 1220 |
| Lattice H diffusivity (m²/s) | $D_L$ | 1.5E-08 | | |
| Partial molar volume of H (m³/mol) | $V_H$ | 2E−06 | | |

## 3. Results & Discussion

Initially, using Eq. (1), the NFS$_\%$ for the three Q & T steels were calculated. Here, it is again mention worthy that the H precharged saturated samples were rapid tested to eliminate the chances of any hydrogen loss. As can be observed from Fig. 2, the NFS$_\%$ in case of 4340 is the lowest while it is maximum for 4140, which can be attributed to the fact that 4340 has the highest strength (both yield and tensile) and hardness [1,16]. However, the Q & T steels 4340L and 4140 have similar hardnesses but significantly different NFS$_\%$ values. Clearly, strength is not the only consideration in the determination of HE susceptibility. This observation can provide insights into the effect of strength on HE susceptibility, which will be explored in this section.

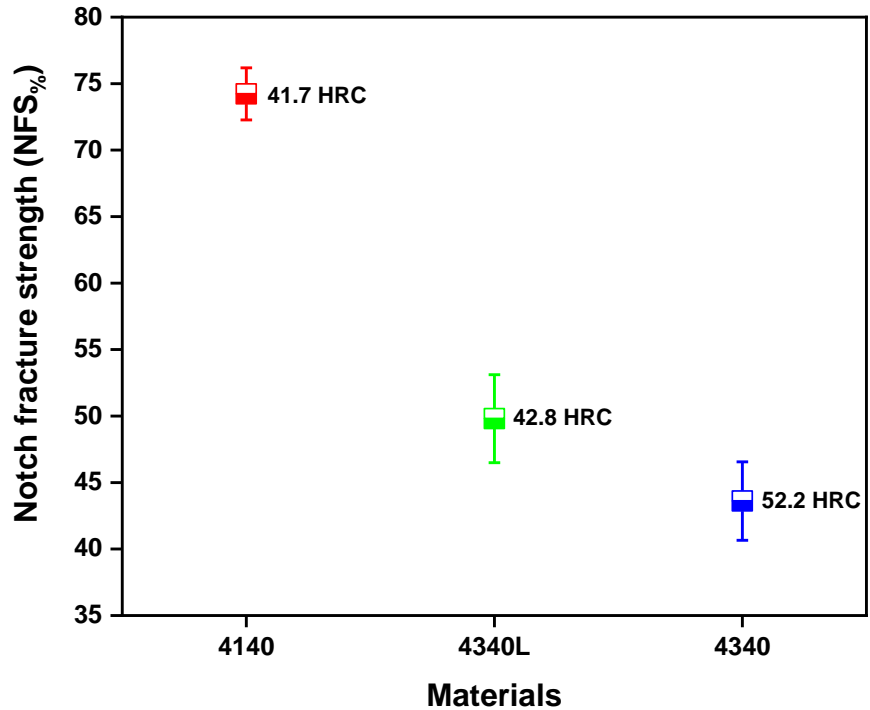

**Fig. 2.** Shows the percent notch fracture strength (NFS$_\%$) for the different materials obtained from the fast fracture test.

The novel fast fracture methodology described in *section 2.2* records the fracture loads and as well as the respective failure time in each case. The average fracture loads, both in the absence (air test) and presence (pre-charged condition) of H, are shown in Fig. 3. The corresponding net section nominal bending stresses, $\sigma_{nom} = 6F_y u/Bz^2$ where $B$ is the thickness (10 mm), $F_y$ is the average fracture load, $u$ is the moment arm (203.2 mm) and z is the uncracked ligament (6.44 mm), were also calculated and given in Fig. 3. In the absence of H, the values of $\sigma_{nom}$ are close to the values reported in [48]. Figure 4 further shows the load versus time plot for the three different steels in the absence of H. In the *absence* of H, it can be seen that 4340L and 4140 have very close average failure loads owing to their *similar* strength/hardness levels, while 4340 has a higher notch fracture strength and takes more time to fracture. However, in the *presence* of H, the behavior changes as can be also observed in Figs. 5 (a), 5 (b) and 5 (c) for 4140, 4340L and 4340 respectively. The average failure loads are close for 4340 and 4340L, but significantly higher in case of 4140. The fast fracture test methodology includes plastic deformation as well as fracture, but the separation of the two is not possible. There is no clear indication in the literature that whether hydrogen significantly influences the yield strength and strain hardening behavior of Q & T steels [43–45,52]. Therefore, as an initial approach, in this study, FEA analyses were carried out to the study the evolution of stress and H concentration distributions, without considering the effect of hydrogen in the constitutive equations.

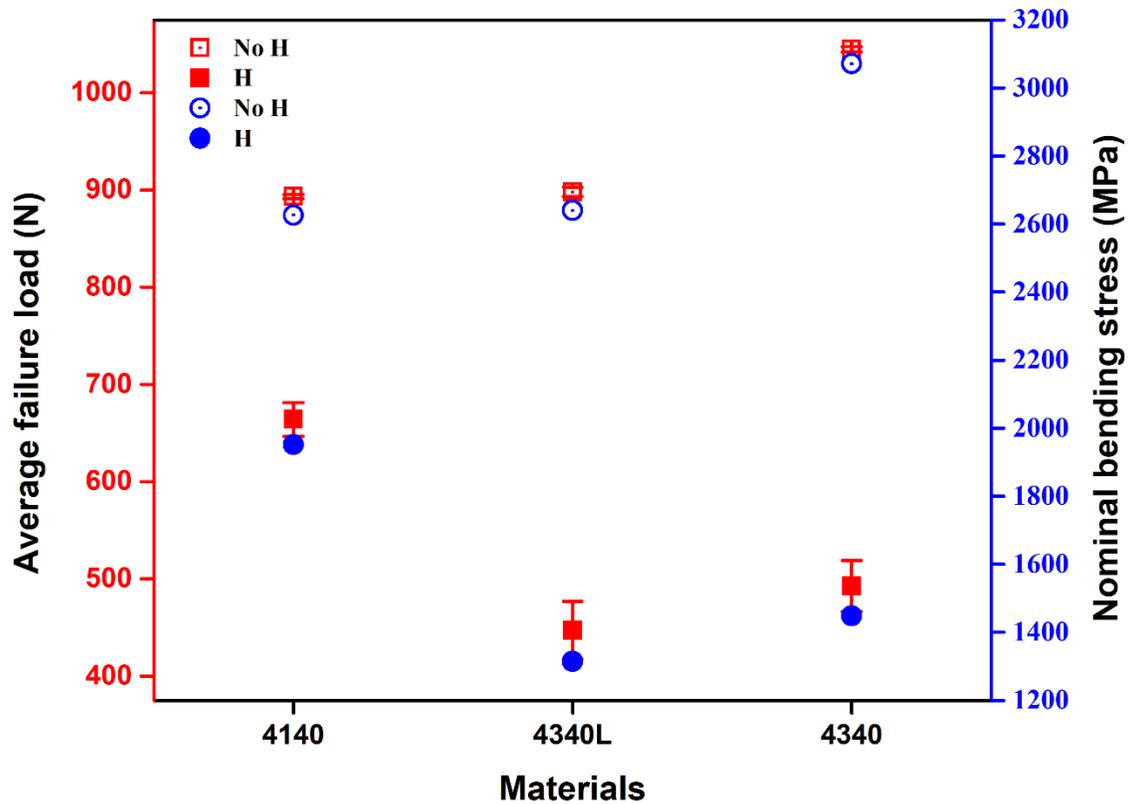

**Fig. 3.** Shows the average failure load and nominal bending stress obtained from the fast fracture test for the different materials both in the absence (open symbols) and presence (closed symbols) of H.

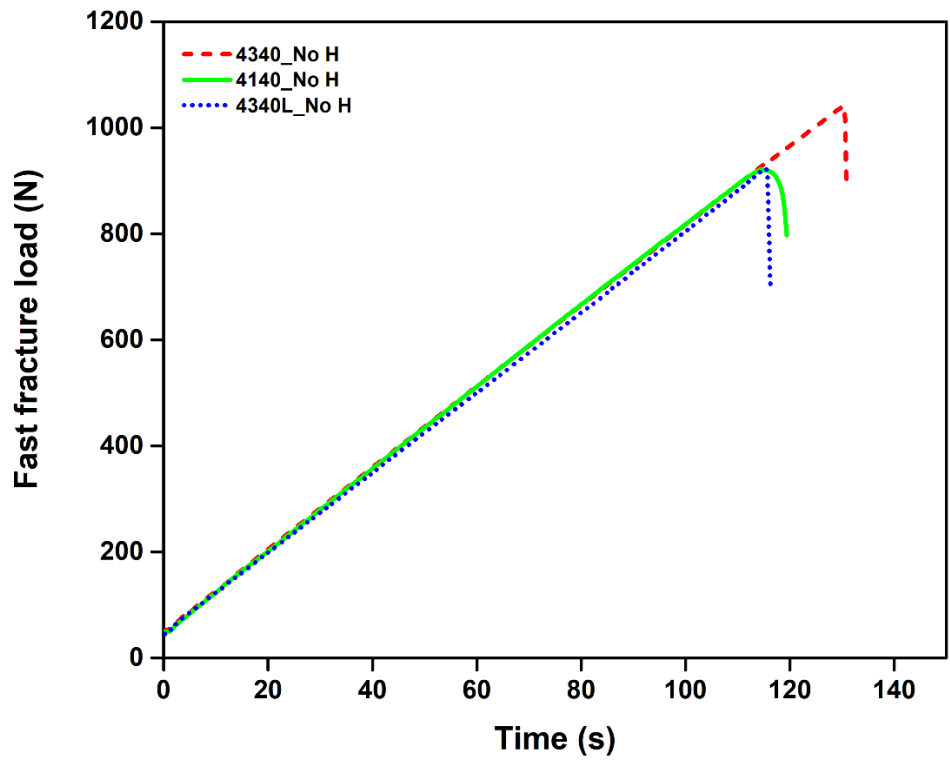

**Fig. 4.** Fast fracture load vs time plot for the three different materials tested in absence of H.

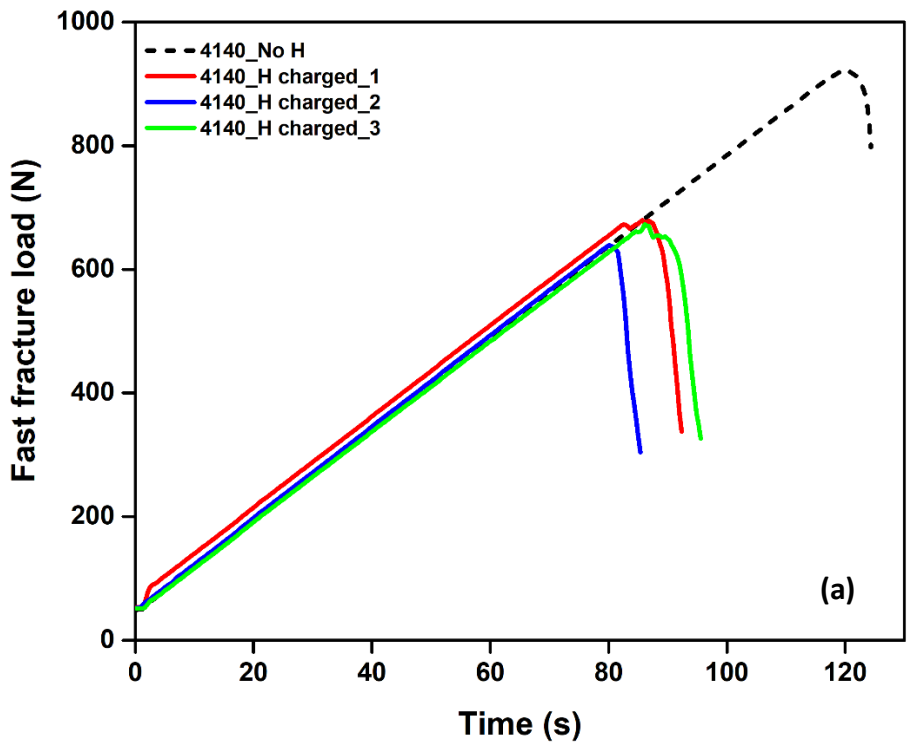

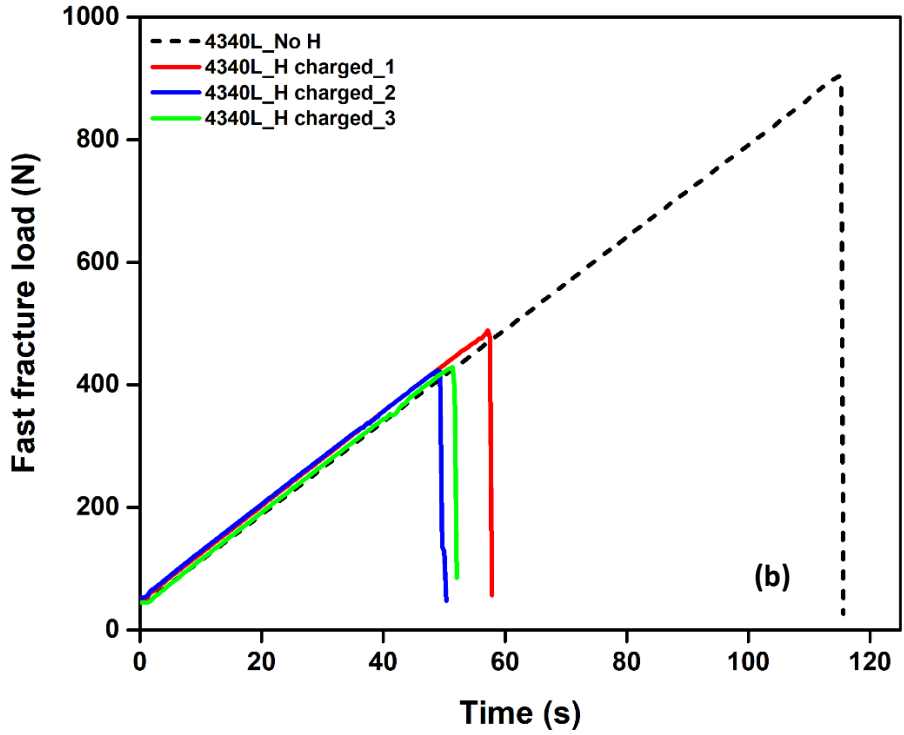

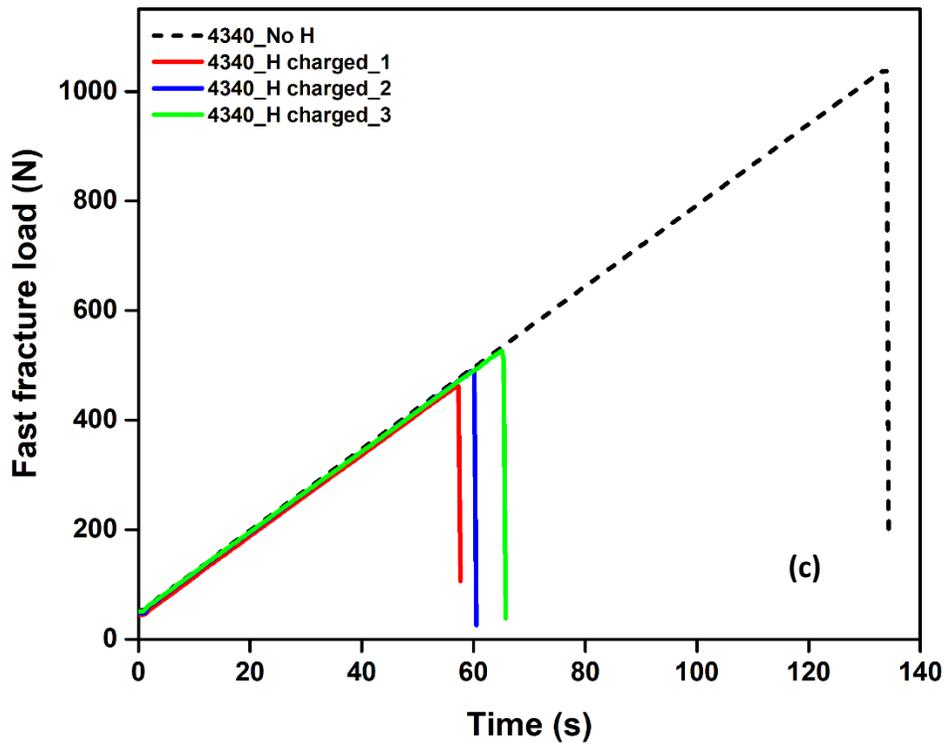

**Fig. 5.** Fast fracture load versus time plot for the three different materials *(a)* 4140 *(b)* 4340L and *(c)* 4340 both in absence (dashed line) and presence (solid line) of H.

The distribution of hydrostatic stress along the notch length at failure are shown in Fig. 6 (a) and 6 (b), in the absence and in the presence of H, respectively. In the *absence* of H (Fig. 6 (a)), it can be observed that, in all cases, the highest values of the hydrostatic stress are not at the notch tip, because they are at the elastic-plastic boundary, and 4340 has the highest value of the three because it has the highest strength. However, *at the notch tip* the lowest hydrostatic stress is exhibited by 4140.

In the *presence* of H (Fig. 6 (b)), it can be noticed that 4140 now has the highest hydrostatic stress (at the elastic-plastic boundary). However, the hydrostatic stress trend *at the notch tip* does not change (i.e. 4140 is the lowest, as shown in the magnified view in Fig. 6 (b)).

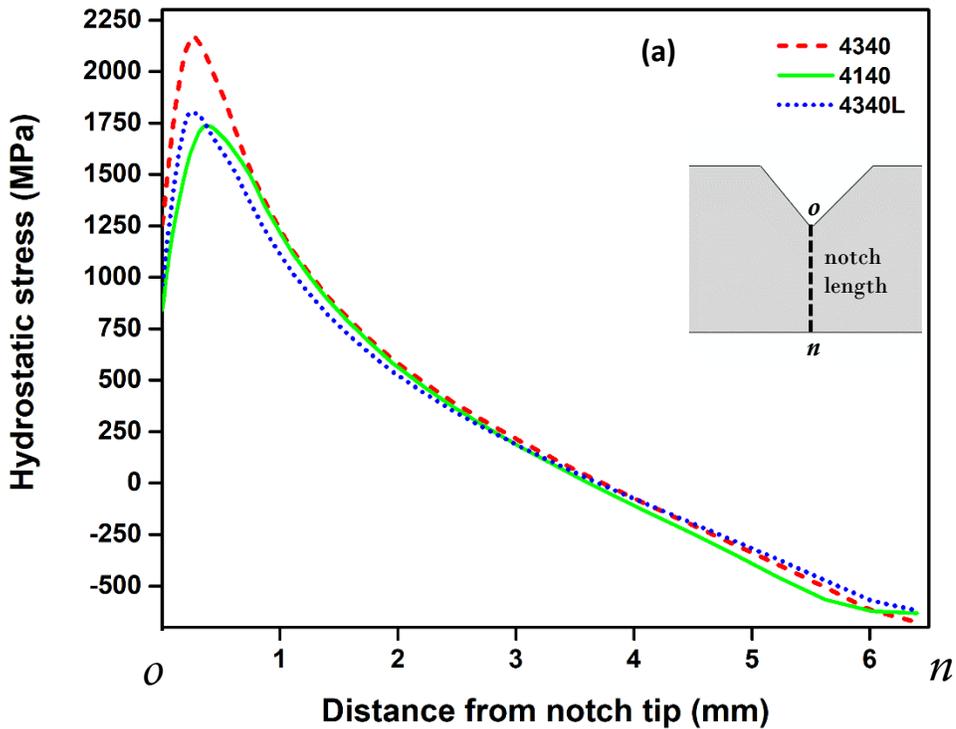

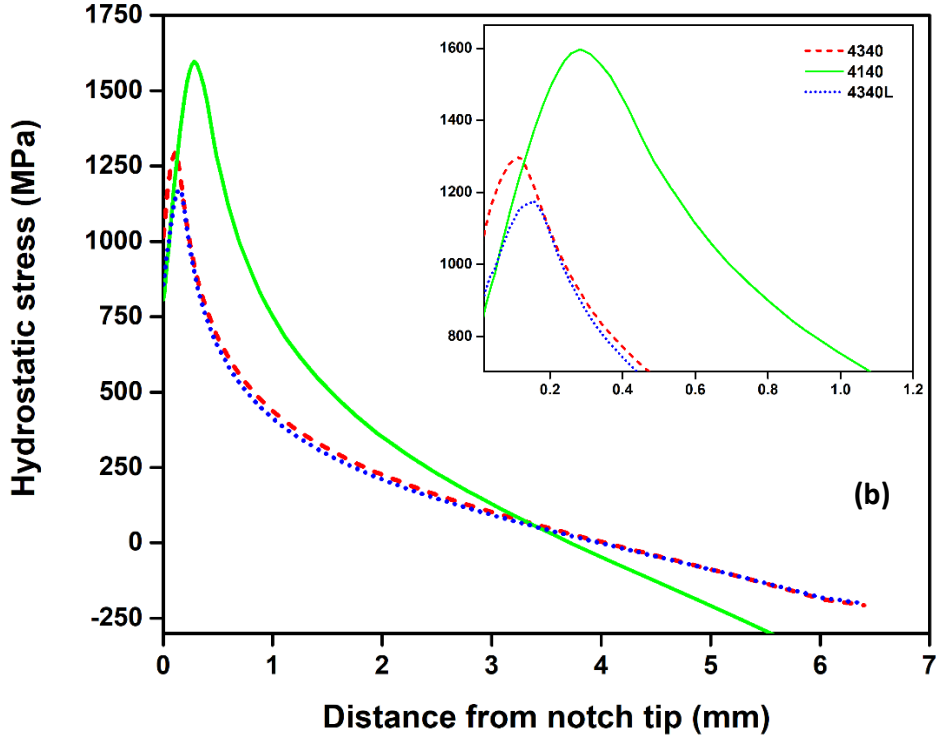

**Fig. 6.** Shows the distribution of hydrostatic stress along the notch length (*on*) at failure in *(a)* absence and *(b)* presence of H respectively, a magnified view of the stresses at the notch tip is shown at the top right corner.

Now, as mentioned in the methodology (*section 2.3.1*), an initial H concentration of 36.305 mol/m$^3$ was considered to be uniformly distributed throughout the geometry. As shown in Fig. 7(a), the *equilibrium* lattice H distribution follows the hydrostatic stress distribution from the stress-coupled H diffusion calculations. Figure 7(a) further validates the stress-coupled H diffusion FEA model. The equilibrium lattice H distribution computed using the FEA model very closely resembles with the calculated values based on the analytical expression, $C_{equilib} = C_0 \exp(\sigma_h V_H / RT)$ [34]. However, the lattice H concentration in *quasi-equilibrium* state reveals a different profile, as can be observed from Fig. 7(b). This hydrogen concentration profile should resemble more closely the experimental condition, because of the rapid nature of loading. At the notch tip, there is a difference of 10 mol/m$^3$ (~1.3 wt. ppm) between 4140 and 4340, while the difference is 8 mol/m$^3$ between 4140 and 4340L. The plastic deformation preceding the failure of 4140 is responsible for the large plastic zone size over which the hydrostatic stress is distributed. It further contributes to the depletion of H from the surrounding region (such as the notch tip), leading to its concentration over the elastoplastic region. This observation can be further manifested from the 2D surface plots. Figure 8 shows both *equilibrium* (Fig. 8: a1-c1) and *quasi-equilibrium* (Fig. 8: a2-c2) distributions of H close to the notch tip at the time of failure. Clearly, under *quasi-equilibrium* condition, the depletion of H from the notch tip and its distribution over the elastoplastic region is more pronounced in 4140 as compared to 4340 and 4340L. On the other hand, the H distribution in the *equilibrium* state exactly follows the hydrostatic stress distribution (Fig. 6 (b) and 7 (a)). This is due to the redistribution of H based on energetics in the surrounding region following the fast depletion of H (Fig. 8: a2-c2). This phenomenon demonstrates the potential role of plasticity and plastic zone size in resisting HE fracture.

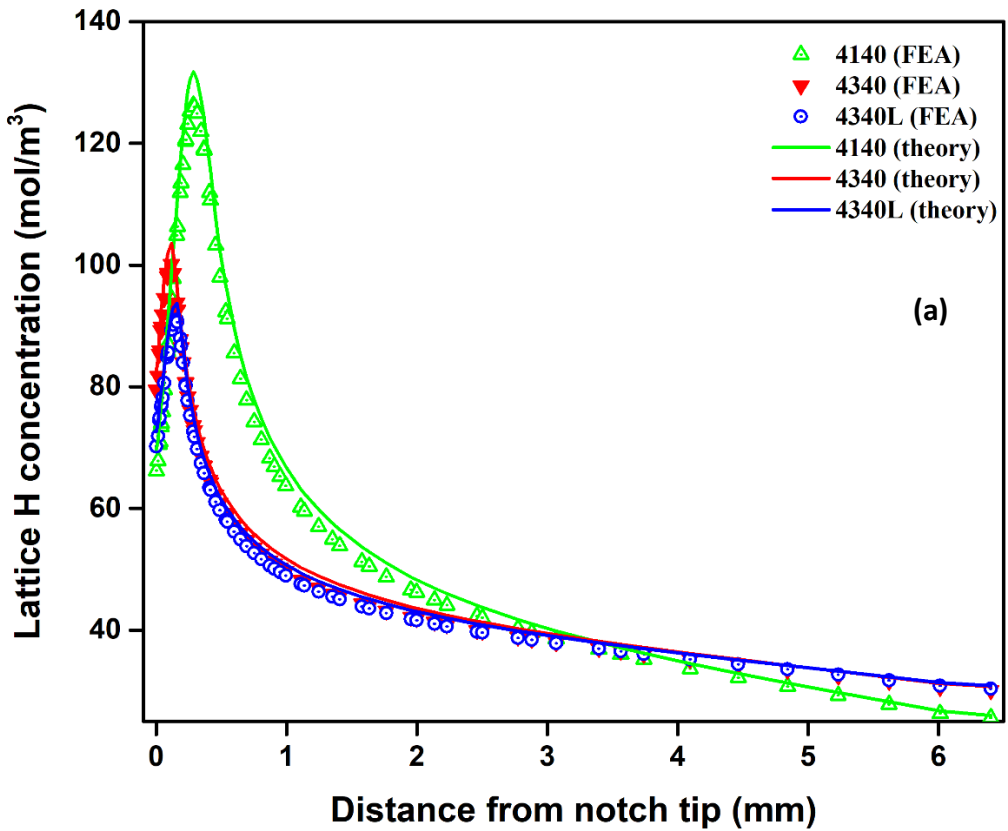

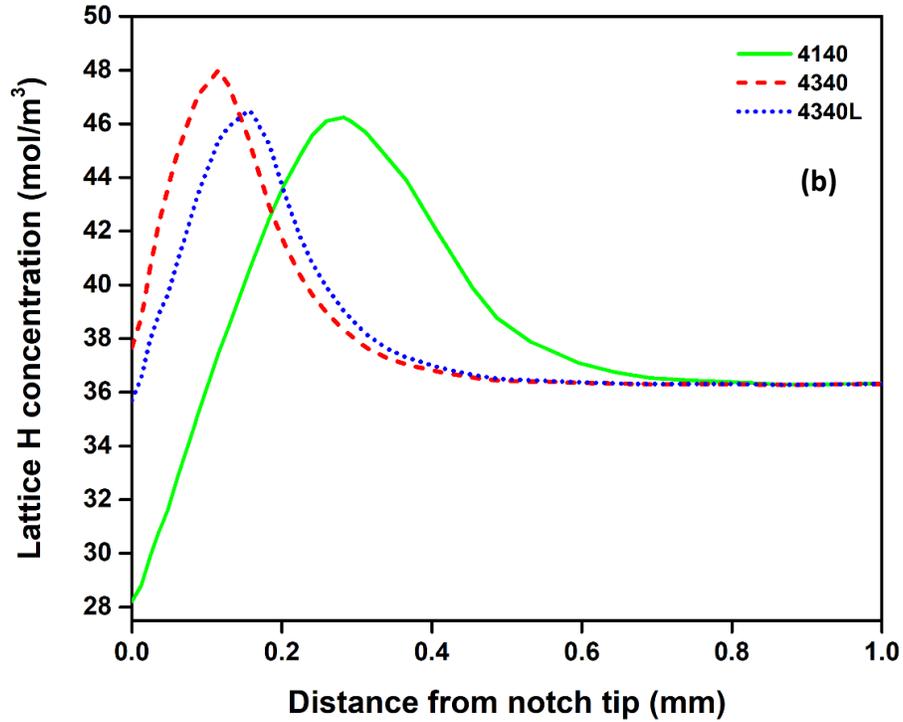

**Fig. 7.** *(a)* Shows a comparison in the distribution of *equilibrium* lattice H concentration along the notch length calculated from theory and FEA simulations, while *(b)* shows the *quasi-equilibrium* lattice H distribution, at failure in the Q & T steels.

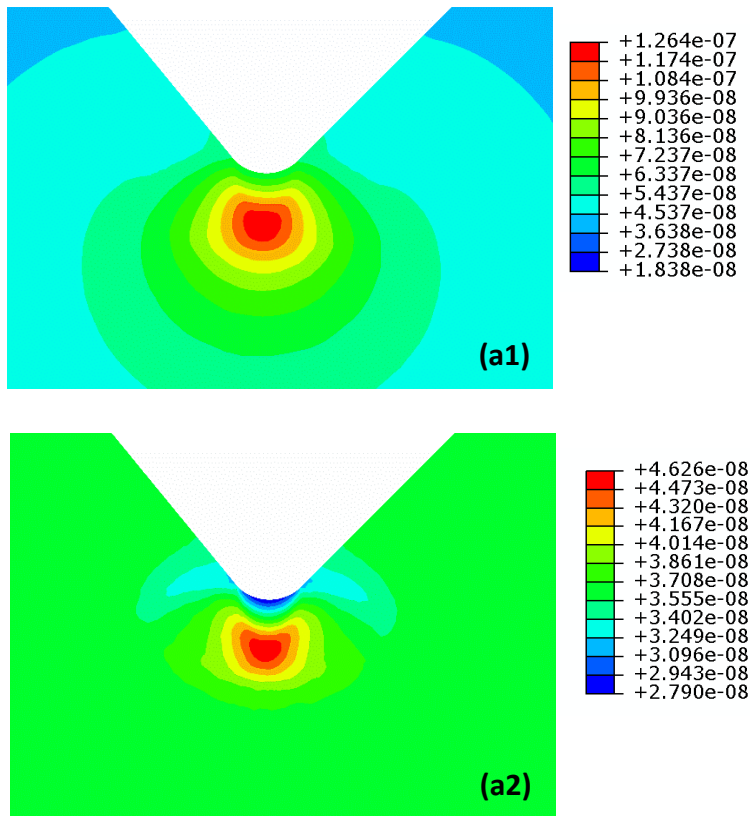

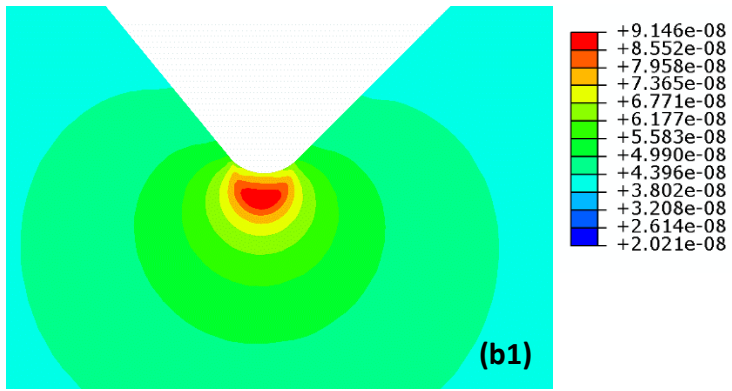

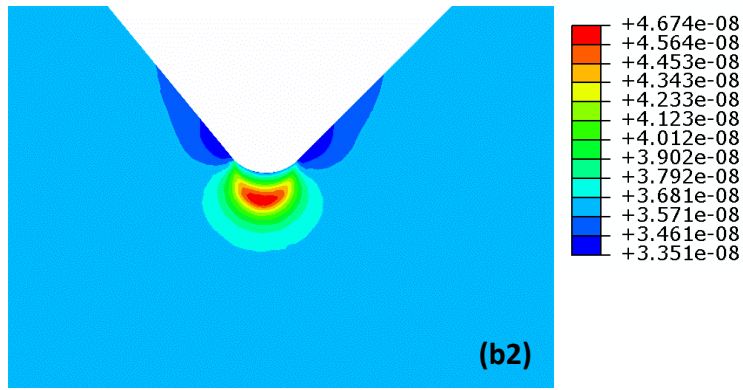

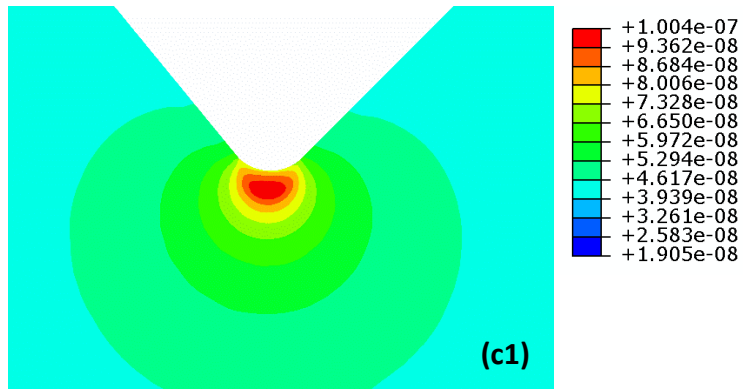

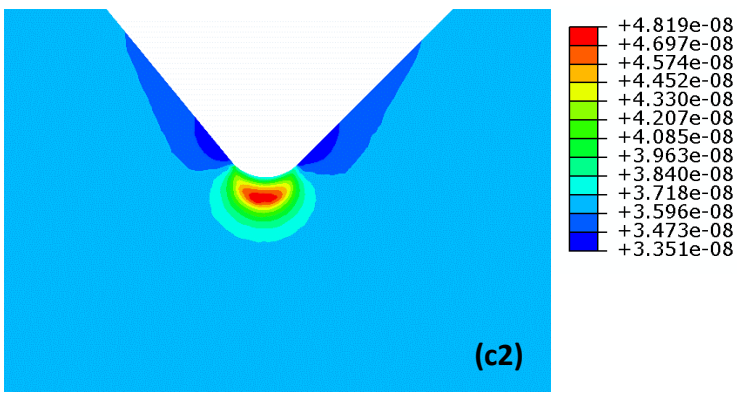

**Fig. 8.** (a1), (b1), (c1) show the 2D surface plots of *equilibrium* lattice H concentration (mol/mm$^3$), while (a2), (b2), (c2) show the *quasi-equilibrium* lattice H concentration (mol/mm$^3$) in 4140, 4340L and 4340 respectively.

However, consideration of only the lattice H concentration to be responsible for HE is not adequate. It is also necessary to question the specific contributions of microstructure in affecting the lattice diffusivity leading to an apparent diffusivity, and consequently change of concentration at the notch tip.

The question of apparent diffusivity has been already considered in [53], where stress coupled diffusion FEA simulations were carried out considering different H diffusivity values for the same AISI 4130 steel. A uniform distribution of hydrogen within the geometry similar to the current experiments was considered as the initial condition. It was concluded from the study, that altering H diffusivity by even three orders of magnitude has a marginal impact on the 'degradation factor' defining the severity of HE failure. It thus suggests that microstructural features impeding H diffusion might not have significant influences on HE in case of internal hydrogen embrittlement (IHE). As a result, the role of microstructure influencing diffusivity, which can be termed as a "global" role, could be neglected to a large extent in this case. Kinetics might lose its engineering importance when long term exposure to a H source is considered, i.e. in some environmental hydrogen embrittlement (EHE) situations. Thus, the role of microstructure altering H diffusion and finally affecting HE, needs more rigorous investigations in regard to initial and boundary conditions, especially concerned to specific engineering applications, and ample literature is already available for a broader understanding on IHE, EHE and H transport kinetics [35,54–57]. Now, due to the rapid nature of loading, chances of long range diffusion of H through the geometry of the sample to the notch root area are also rare. However, an additional concentration build up in the proximity of highest stress concentration site is possible, due to local diffusion of H from the neighborhood of the notch tip area, as was illustrated in Fig. 8, from the stress coupled H diffusion simulations. Consequently, the above discussion reduces to the second question related to the role of local microstructural evolution at the notch tip, influencing the micromechanics of HE failure.

In a recent work [52] on a tempered martensite steel, the individual roles of diffusible and deeply trapped hydrogen on HE failure was identified, however, the roles of lattice H and reversibly trapped H (e.g. H trapped by dislocations etc.) were not separated. The lattice and reversibly trapped H were treated as one i.e. diffusible H. But, as far as the contribution of diffusible H is concerned, it is rather detrimental and mainly related to the quasi-cleavage fracture observed in that study. Hence, it has been assumed in this study that the deeply trapped H concentration might have a second order effect and thus, has not been evaluated. More emphasis has been put into the calculation and discussion concerning the role of lattice H and reversible trapped H in connection to the observations made in this study. Now, as per the FEA analyses, since all the three materials undergo plastic deformation before failure both in presence and absence of H; there is a development of plastic strain in the notch tip region. Figure 9 shows the plastic strain distribution along the notch length of 4340, 4140 and 4340L at the time of failure, in the *presence* of H. It can be observed that the plastic strain as well as the plastic zone size is highest in case of 4140 followed by 4340L and 4340. Now, using plastic strain, the trap density associated with dislocation density can be evaluated using Eq. (9 & 10). Based on which the trapped H concentration can be further estimated according to Eq. (8), using the *quasi-equilibrium* lattice H concentration (from Fig. 7 (b)). The trapped H concentrations due to plastic strain (or dislocations) along the notch length are shown in Fig 10. Therefore, hydrogen moving from lattice sites to trap sites will be highest in case of 4140 inside the plastic zone. Moreover, it can be further noticed from Fig. 7, lattice H concentration in case of 4140 remains lower than 4340 by approximately 10 mol/m$^3$ (~1.3 wt. ppm) at the notch tip. This observation suggests that the lower lattice H concentration profile in the near notch region in case of 4140 contributes to its higher fracture load as opposed to 4340. This phenomenon can be only effective if extensive plastic deformation occurs. However, limitation of plasticity and consequent reduction of plastic zone size in case of in presence of H are reported in several papers for iron and its alloys [58–62]. Nevertheless, even if, there is reduction in plastic zone size in presence of H, 4140 will have advantage over the other materials and thus, it qualitatively explains its lowest susceptibility. Hydrogen trapping by dislocations could be pertinent, depending on the nature of H trapping by dislocations. The trapped H concentrations were estimated by considering 18 kJ/mol binding energy in case of 4340 and 4340L [48] and 30 kJ/mol in case of 4140 [63] respectively. The higher plastic strain and the higher trap binding energy associated with dislocations may result in more effective trapping of H in case of 4140. Again, the occurrence of H deep trapping (48.24 kJ/mol-57.9 kJ/mol) in the cores of the dislocations associated with plastic strain, was reported in a martensitic steel [64]. Therefore, whether it is reversible or irreversible H trapping, the outcome will work in favor of 4140 in resisting HE failure.

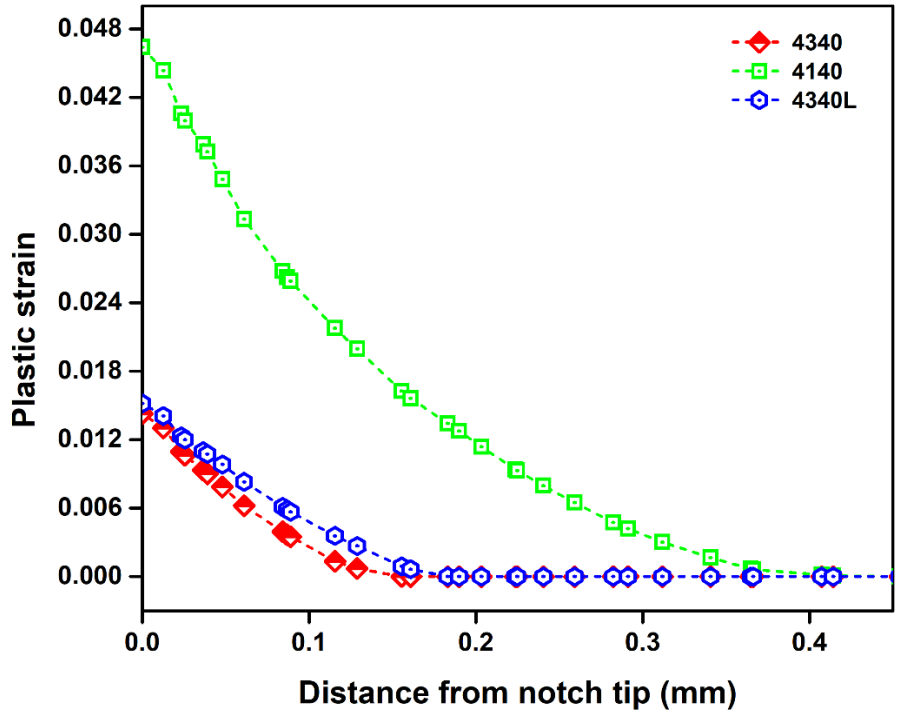

**Fig. 9.** Shows the plastic strain distribution along the notch length at the time of failure in the three different Q & T steels.

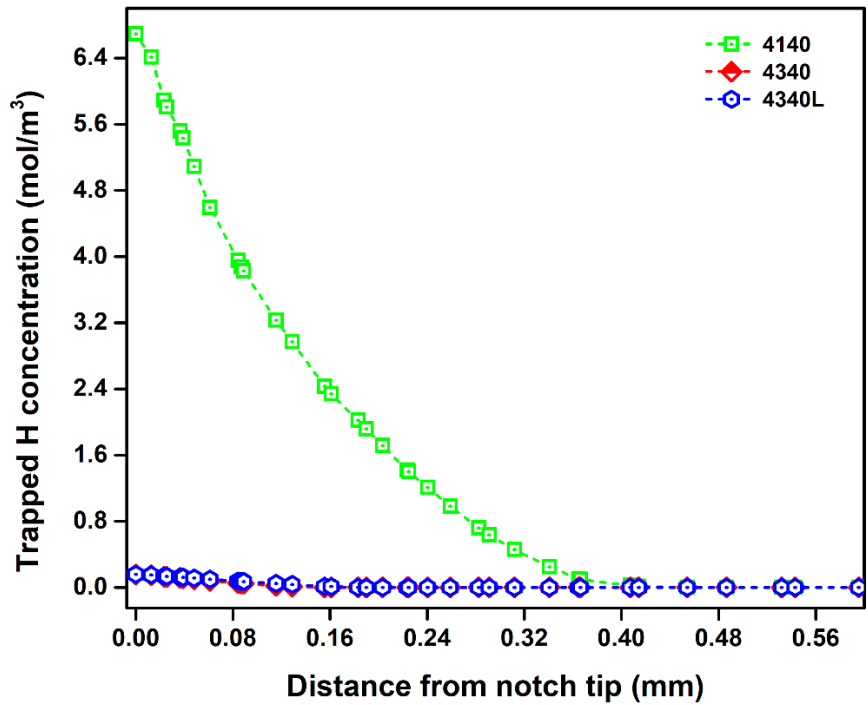

**Fig. 10.** Shows the distribution of equilibrium trapped H concentration along the notch length at the time of failure in the three different Q & T steels.

In order to understand the behavior exhibited by 4340L and 4340 in Fig. 3, where 4340L and 4340 fail approximately at the same load in presence of H, a comparative study is required. The hydrostatic stress distributions for all three materials at the failure time of 4340L (i.e. 55s, the shortest time) are shown in Fig. 11 to elucidate the state of stress among the three steels. As can be observed, in a magnified view in Fig. 11, the maximum hydrostatic stress at the elastic-plastic boundary of the three materials are very close. The behavior observed in Fig. 11 in terms of similar maximum hydrostatic stress (at the elastic-plastic boundary), could be an outcome of small scale yielding undergone by the materials, and differences in strain hardening behavior among them around that time. Nevertheless, the hydrostatic stresses are different at the notch tip, being highest in case of 4340. The distribution of lattice H concentration along the notch length in quasi-equilibrium is analogous to the hydrostatic stress distribution, as can be seen in Fig. 12. The plastic strain distributions along the notch length in the steels at 55s are shown in Fig. 13. The 2d surface plots of *quasi-equilibrium* lattice H distribution in the notch area in the three materials are also shown in Fig. 14.

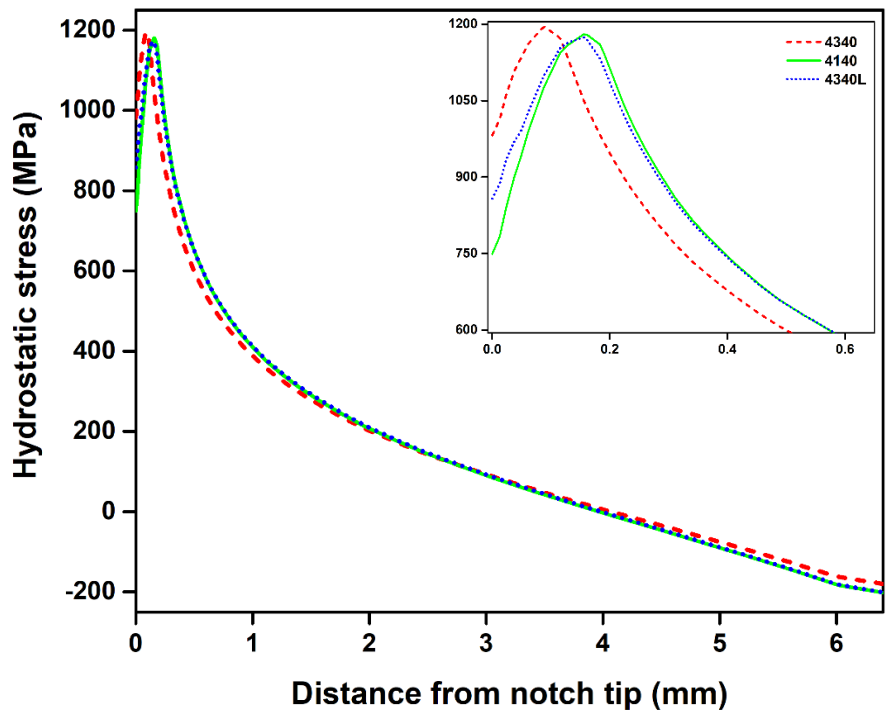

**Fig. 11.** Shows the hydrostatic stress distribution profile along the notch length at 55s in the Q & T steels, a magnified view of the stresses at the notch tip is shown at the top right corner.

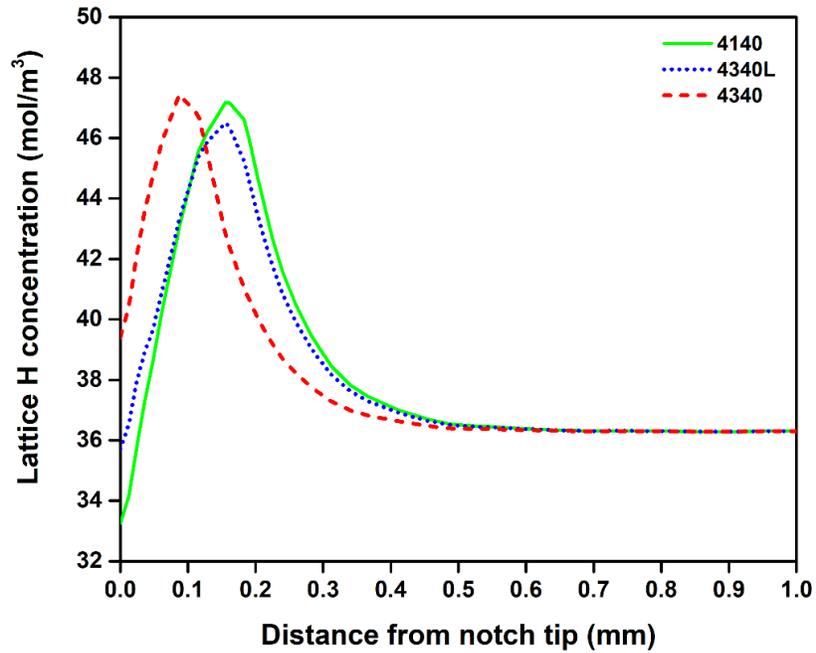

**Fig. 12.** Shows the *quasi-equilibrium* lattice H profile along the notch length at 55s following the hydrostatic stress distribution from Fig. 11.

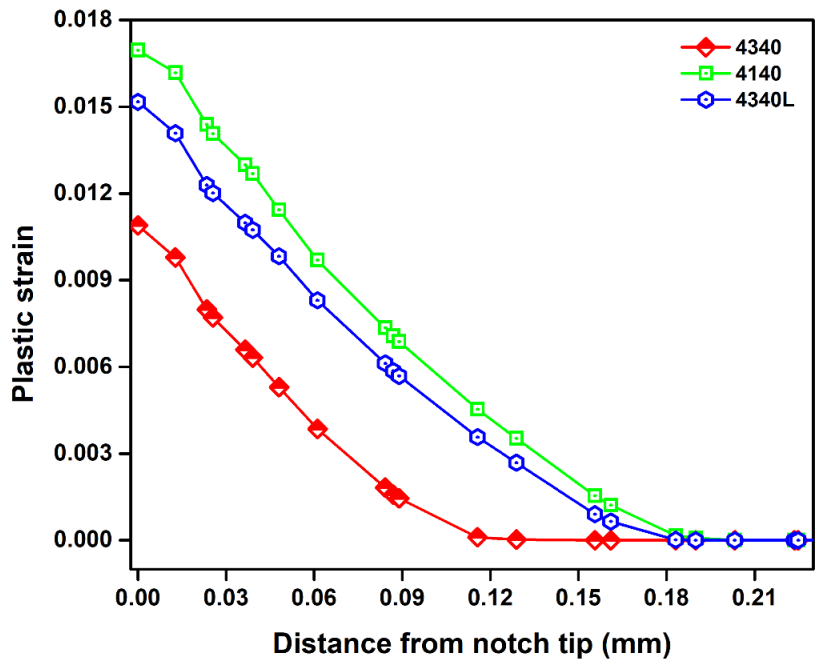

**Fig. 13.** Shows the plastic strain distribution along the notch length at 55s.

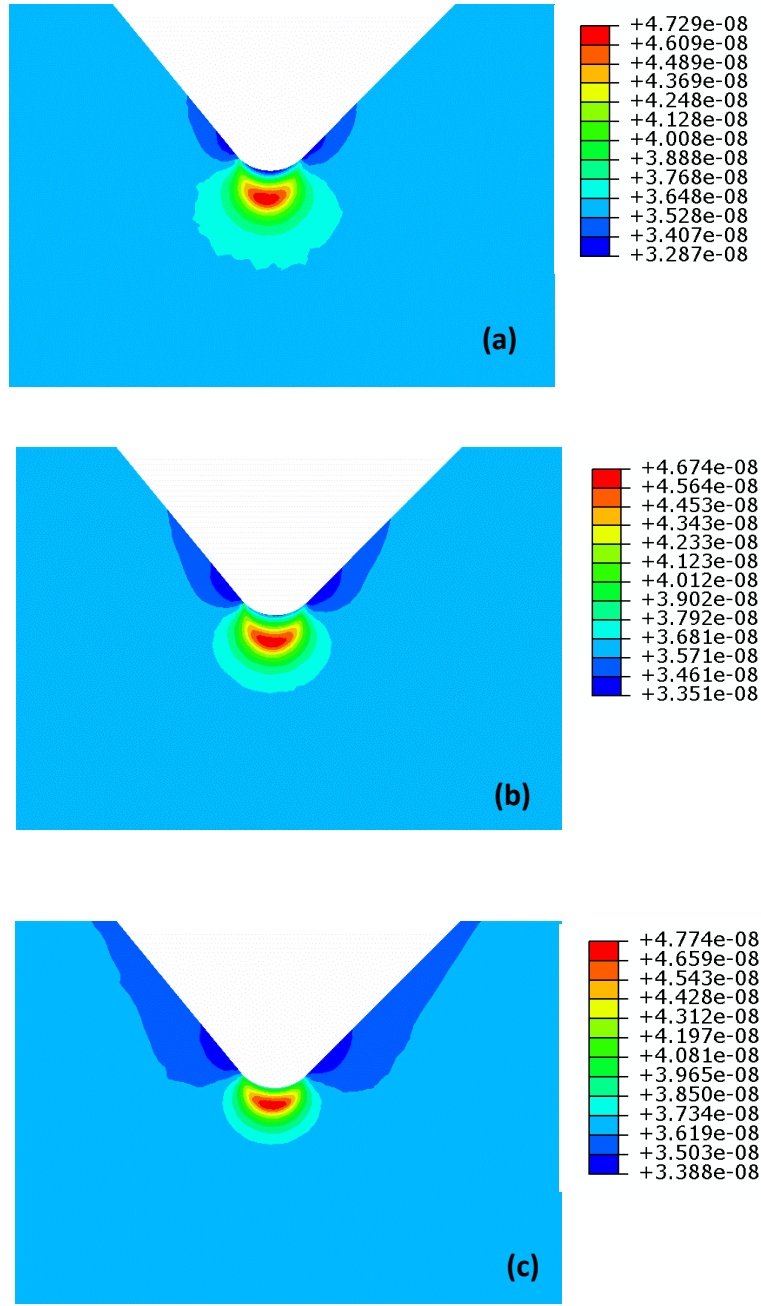

**Fig. 14.** 2D surface plots showing *quasi-equilibrium* lattice H concentration (mol/mm$^3$) distribution in *(a)* 4140, *(b)* 4340L and *(c)* 4340 respectively.

Clearly, 4140 has highest plastic strain and largest plastic zone size followed by 4340L and 4340. Another striking observation is the more concentrated H distribution field in 4340L and 4340 as compared to 4140 (Fig. 14). Therefore, it is not only the stress and concentration profile, but also field distribution of the same could be of particular importance. Dislocation induced H trapping can also occur in 4340L, however, no significant differences in trapped H concentration can be observed between 4340L and 4340 (Fig. 15). It is further observed from Fig. 12, there are almost no differences in lattice H concentration at the elastic-plastic boundary among the three steels. However, differences can be observed at the notch tip, where 4340L has approximately 5 mol/m$^3$ less amount of lattice H concentration than 4340. In spite of 4340 having a higher lattice H concentration and higher hardness, 4340 and 4340L fail nearly at the same time and load in presence of H. This observation essentially leads to question the role of strength influencing HE failure, because 4340 and 4340L have difference in hardness by approximately 10 HRC and noticeably different strength levels, yet their failure conditions are remarkably close. This observation implies that in this range of strength level, there is another first order effect, which could be the critical H concentration; this can be defined as the concentration at which the H induced cracking triggers macroscopic fracture of the material. Earlier, the concept and evaluation methodology of critical H concentration have been reported by several researchers [65–67,25] with a consensus that there is a ductile to brittle transition, particularly in steels, when a certain H concentration level is reached. Hence, based on the concept of critical H concentration, 4340L has the same or lower level of critical H concentration requirement than 4340. Thus, the performance of in 4340L a H environment is not more promising than 4340, even though, the load drops i.e. NFS$_%$ of 4340L is less as compared to 4340. However, an explanation based on the critical H concentration without suitable validation is rather intuitive than quantitative. Therefore, a mechanistic description has been provided in the following section.

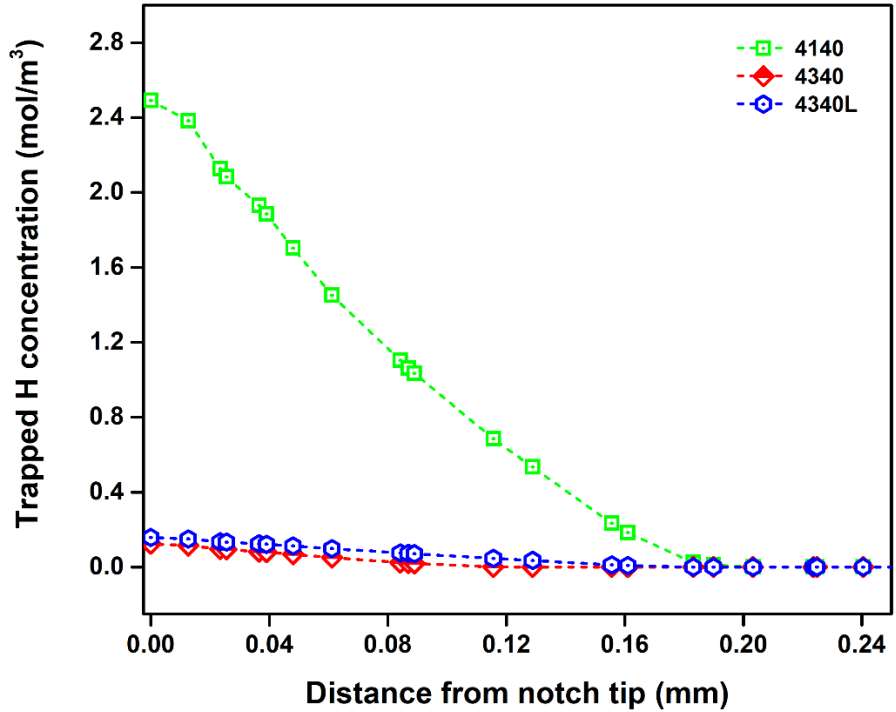

**Fig. 15.** Shows the trapped H concentration profile along the notch length at 55s.

## 3. 1. A mechanistic description

In this section, an extended discussion based on analytical and fractography techniques is presented to explain the similar behavior of 4340 and 4340L in the presence of H. An effort has been made to understand the plastic deformation of the materials in the presence of H. In this context, two aspects are important: a) dislocation emission from a crack tip and b) dislocation interaction resulting in strain hardening.

There could be two possible aspects while considering the effect of hydrogen in the near crack tip region. Hydrogen can either limit overall dislocation emission from a crack tip or can render lower fracture (Griffith) energy along the plane where the crack tip is situated, resulting in brittle cleavage failure before adequate dislocation emissions can occur. However, considerable contention remains in the literature regarding the occurrence of the first aspect. As a result, attention has been mainly paid to the latter aspect in this study.

In the case of mode I loading, based on the results presented in [68], the minimum critical applied stress intensity factor (SIF) necessary for dislocation emission from the crack tip, $K_I^e$, occurring at the slip plane inclined at angle θ = 70.53° with respect to the horizontal axis can be evaluated as 0.904 MPam$^{0.5}$. For simplicity, $K_I^e$ was calculated considering the parameters of bcc iron. Now, the local SIF for dislocation emission at the crack tip is either reduced or enhanced from, due to the shielding or anti-shielding effects of all the dislocations given by Eq. (16) [69,70]:

$$K_{tip} = K_I - K_d \tag{16}$$

where $K_I$ is the remotely applied SIF, and $K_d$ encompasses the shielding/anti-shielding effect, that be evaluated from the Atkinson and Clements anisotropy solution of crack-tip dislocation interactions provided in [71] using complex functions developed by Stroh [72]. The remotely applied SIF, $K_I$ can be further evaluated for a SE(B) specimen in pure bending from the expression given in [73] as:

$$K_I = \sigma\sqrt{\pi a} f\left(\frac{a}{W}\right) \tag{17}$$

In Eq. (17), σ is the gross nominal stress and $f\left(\frac{a}{W}\right) \approx 1.18$, is a function of the specimen and crack geometry. Again, based on the dislocation dynamics simulations carried out by Gerberich and coworkers in case of Mo single crystals, Fe-3wt%. Si [28,74], it was observed that the local SIF, $K_{tip}$ increases considerably both with the increase in yield strength and $K_I$. In their simulations, the forces on individual dislocations are carried out to satisfy both the quasi-equilibrium condition and the tip-emission condition after $K_I$, and the lattice friction strength at σ$_{ys}$/2 were specified. The shielding effect is further evaluated in terms of $K_{tip}$ on the basis of the formulation given in [74]. The detailed algorithm involving this Mode I simulation has been documented in [26–28]. An expression for the local SIF, $K_{tip}$ based on the continuum solution of Lin, Thomson and Weertman for dislocation pile up near a crack tip given in [28] as:

$$K_{tip} = \chi \sigma_{ys} \sqrt{\psi} \left[ ln\left(\frac{4r_p}{\psi}\right) + 4/3 \right] \tag{18}$$

where $\chi = 3/\pi \sqrt{2/\pi}$, $r_p$ is the plastic zone size given by $K_I^2/3\pi\sigma_{ys}^2$. $\psi$ is the length of the *dislocation free zone* (DFZ). The existence of DFZ has been confirmed and estimated through TEM observations and theoretical calculations by Gerberich and coworkers [26–28]. At high $K_I$ values, these zones get indistinguishably small and are referred to as tip-emission adjusted zones. Now, $\psi$ can be mathematically expressed as [28]:

$$\psi = \frac{\alpha}{\sigma_{ys} \ln(\beta K_I)} \tag{19}$$

In Eq. (19), $\alpha$ and $\beta$ are constants estimated from the dislocation dynamic simulations as 1.06E-05 MPa.m and 20 MPa$^{-1}$m$^{-0.5}$ respectively in case of Fe-3wt%. Si. These values will be used in the current paper considering the similarity of the materials.

The Griffith value for cleavage fracture along a particular plane *(hkl)* such that the crack tip lies in the *hkl* plane under plain strain condition can be further calculated from Eq. (20) [28] as:

$$k_{G0} = \sqrt{\frac{2E_{[hkl]}\gamma_{s0}}{1-\nu^2}} \tag{20}$$

where $E_{[hkl]}$ is the modulus along the hkl plane and $\gamma_{s0}$ is the surface energy per unit area. The $k_G$ values (in the absence of H) for α-Fe (110), α-Fe (100) and α-Fe (111) planes are also evaluated.

Again, the change in surface energy with H coverage factor, $\Gamma$ for a particular plane is given by [75]:

$$\gamma_{sH}(\Gamma) = (1 - 1.0467\Gamma + 0.1687\Gamma^2)\gamma_{s0} \qquad (21)$$

such that $\Gamma = C_L/C_L \exp(-\Delta g_b^0/RT)$, where $\Delta g_b^0$ is the free energy difference (30 kJ/mol) between the adsorbed and bulk standard states based on Langmuir–McLean isotherm [76]. Figure 16 shows the change in $K_{tip}$ and plastic zone size ($r_p$) with $K_I$. The estimated $K_I^e$ as well as $k_{G0}$ for Fe (111) plane is also shown in the same figure. Now, using Eqs. (20) and (21), the $k_{GH}$ values in presence of lattice H concentration (from Fig. 12) were calculated for the three different Fe planes and tabulated in Table 4. It can be observed that there is a drop in the Griffith value for cleavage fracture in presence of H and lies well below the $K_{tip}$ value. However, the theoretically estimated $k_{GH}$ values could be a slight oversimplification of the real phenomenon, because the choice of parameters is based on pure bcc iron, and the selection of $\Delta g_b^0$ can have an effect. But the calculations also imply that in the presence of H, nearly equal $k_{GH}$ could exist for 4340 and 4340L (Table 4). This estimation is not counter intuitive, since the chemistries of both the steels are the same. Now, as calculated from Eq. (17), 4340L and 4340 undergo HE failure at a $K_I$ value of 34 MPa m$^{0.5}$ and 37.5 MPa m$^{0.5}$ respectively. Figure 17 shows a magnified view of Fig. 16 in the lower $K_I$ range. It can be observed from Fig. 17, 4340 and 4340L have very close $K_{tip}$ values for a plastic zone size of around 65 microns and below, suggesting similar dislocation emission behaviors of the materials. Thus, in the presence of H, both the materials can lose the ability to emit dislocations in a similar manner for such small plastic zone sizes. The question following this observation is whether the possibility of such small plastic zone size formation and constrained plastic behavior exists in these steels. To answer this question, Fig. 17 can be recalled again. The plastic zone sizes are close to a certain extent at low $K_I$ values, especially at the failure condition of $K_I$ = 34 and 37.5 MPam$^{0.5}$, i.e. approximately 76 and 65 microns for 4340L and 4340 respectively. However, these calculations do not capture the impact of H interacting with dislocations resulting in a reduction of the plastic zone size. But, as already mentioned before, the limitation of plasticity in presence of H has already been reported. One such example is the blockage of dislocation emissions from a crack tip in the presence of high local H concentration, leading to the confinement of plastic zone size during in-situ fracture experiments of FeAl micro-cantilever specimens [60]. Hydrogen induced sharp crack formations accompanied by quasi-cleavage fracture with reduced crack-growth resistance in α-iron during elasto-plastic fracture toughness tests were also observed [62]. Therefore, the above phenomenon is also expected to be observed in our experiments. This expectation has been visually evident from fractographic observations, where Fig. 18 and 19 shows the fracture surfaces from the notch tip region in case of 4340L and 4340 respectively (A more extensive presentation will follow in the upcoming section 3.2). It can be observed that the plastic zone sizes are vanishingly small in both the steels, because of the presence of predominant intergranular features right at the notch tip, despite of the differences in strength. Hence, the above discussion also rationalizes our observation from the fast fracture experiments (Fig. 3). However, it is still relevant to question why the plastic zone size and plasticity decreases drastically in presence of H. In order to resolve this question, further understanding would be imperative on how H interacts with dislocations influencing dislocation dynamics and ultimately work hardening, which is beyond the scope of the current study. Nevertheless, the effect of H on work hardening exponent as compared to yield strength has already been reported based on FEA calculations, particularly for steels [77]. It was observed that H reduces the work hardening exponent, consequently reducing plasticity and plastic zone sizes. In addition to experiments and continuum calculations, molecular dynamics (MD) simulations also support this observation in the decrease of plastic zone size in presence of H [78], where an empirical relation was derived between plastic zone size and H concentration in case of X100 steel. Therefore, for better quantitative predictions through theoretical modeling, understanding H-dislocations interactions are of paramount importance.

It should be also noted that linear elastic fracture mechanics (LEFM) with small scale yielding framework has been assumed to conduct all the calculations in this section. As discussed above, this assumption is particularly valid while treating fracture in the *presence of H*, where the plastic zone sizes reduces significantly. The same framework might not bring appropriate justification to the observations made in the *absence of H*, due to substantial deviation from LEFM.

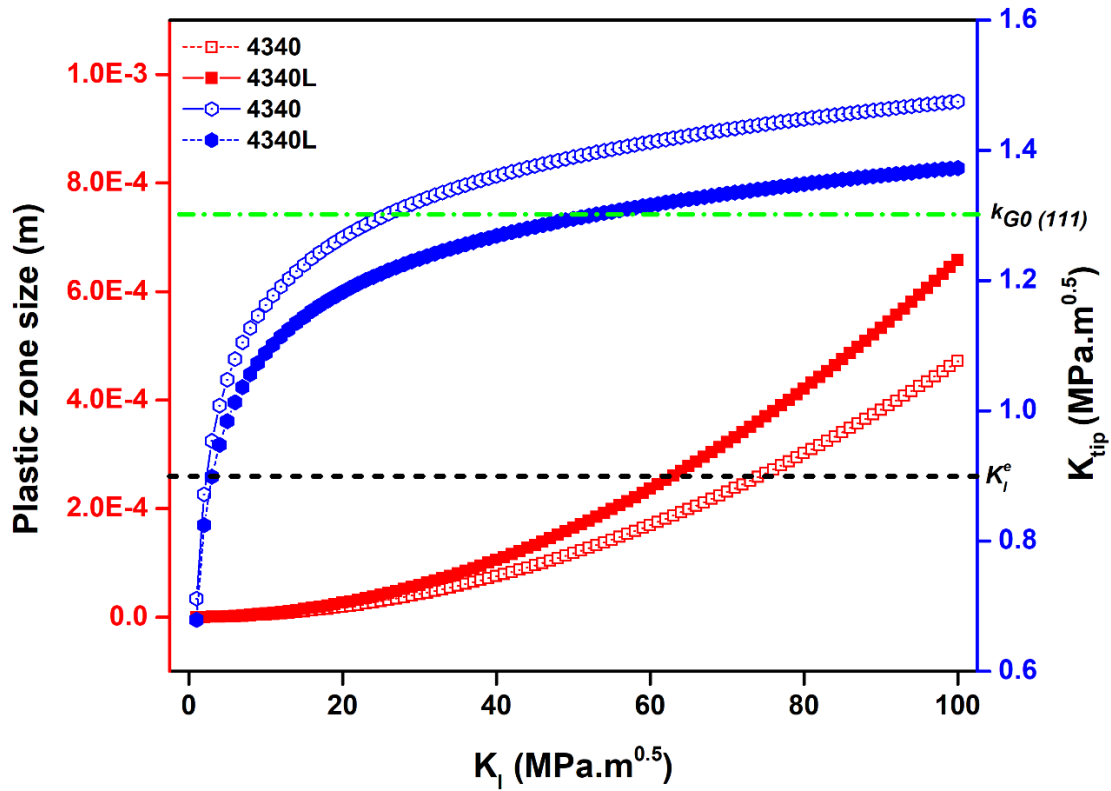

**Fig. 16.** Shows the change in plastic zone size (m) and local SIF ($K_{tip}$) with the increase in the remote applied SIF $K_I$ upto a value of 100 MPam$^{0.5}$ for 4340 and 4340L.

### Table 4

Surface energies and Griffith fracture values for different α-Fe planes in 4340 and 4340L with and without H

| Fe Planes | $E_{[hkl]}$ (GPa) [79] | $\gamma_{s0}$ (J/m$^2$) [80] | $k_{G0}$ (MPam$^{0.5}$) | $\gamma_{sH}$ (J/m$^2$) 4340L | $k_{GH}$ (MPam$^{0.5}$) 4340L | $\gamma_{sH}$ (J/m$^2$) 4340 | $k_{GH}$ (MPam$^{0.5}$) 4340 |
|---|---|---|---|---|---|---|---|
| 110 | 220.5 | 2.47 | 1.094 | 0.3013 | 0.4 | 0.3013 | 0.4 |
| 100 | 129 | 2.56 | 0.852 | 0.3123 | 0.3 | 0.3123 | 0.3 |
| 111 | 276 | 2.78 | 1.3 | 0.3392 | 0.454 | 0.3392 | 0.454 |

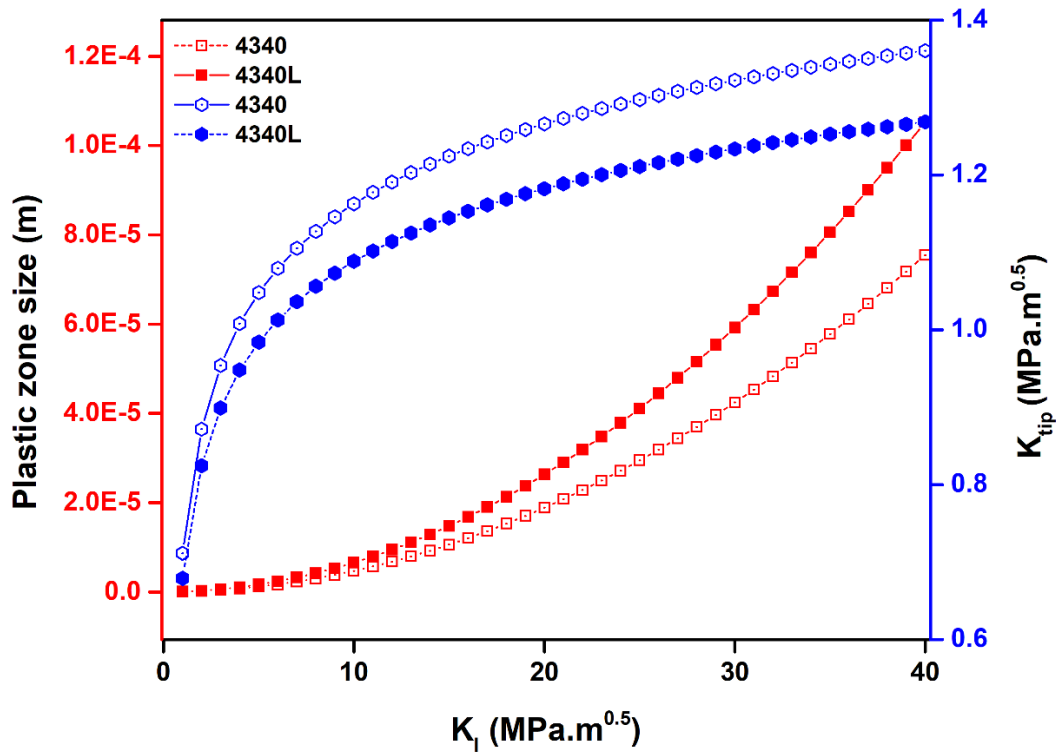

**Fig. 17.** A magnified view of Fig. 16 in the lower range of $K_I$ showing the plastic zone sizes and $K_{tip}$ values near failure condition of the materials (in presence of H), i.e. $K_I$ = 34-38 MPam$^{0.5}$.

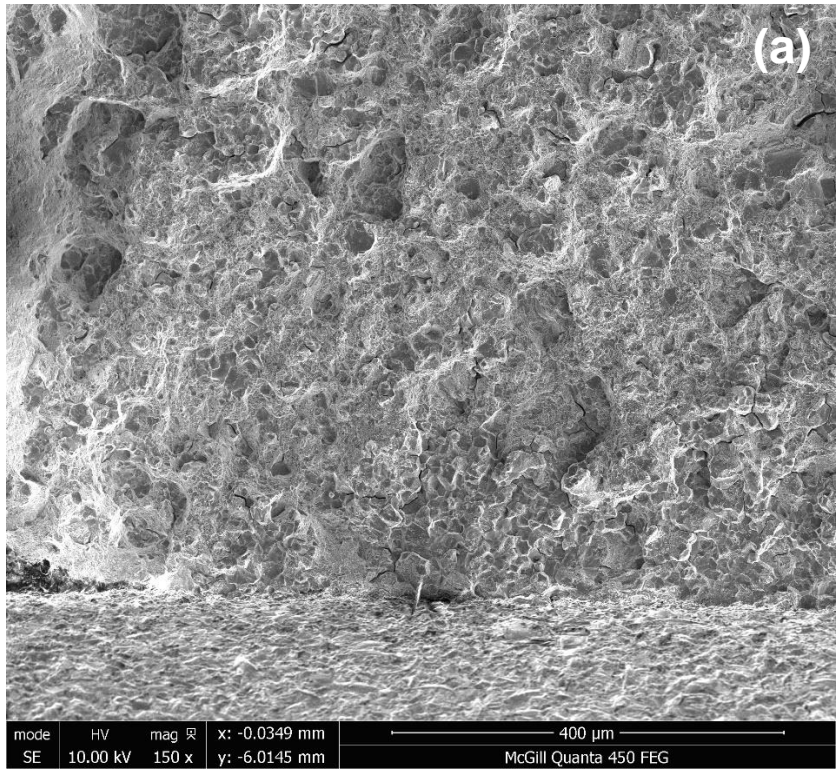

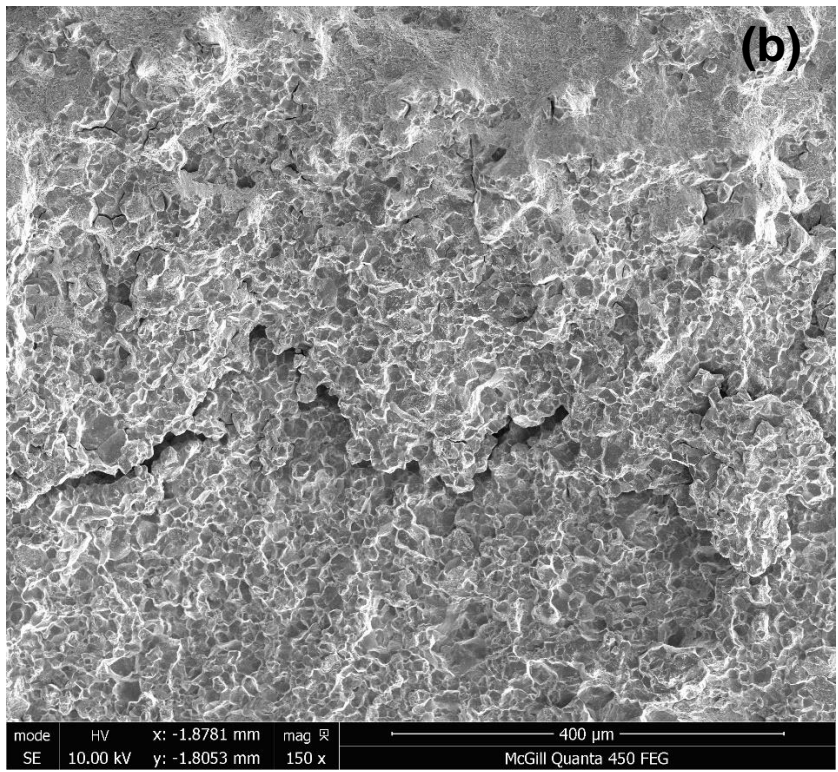

**Fig. 18.** *(a)* and *(b)* shows the fracture surfaces from the near notch tip region in 4340L at 150x for a large field of view.

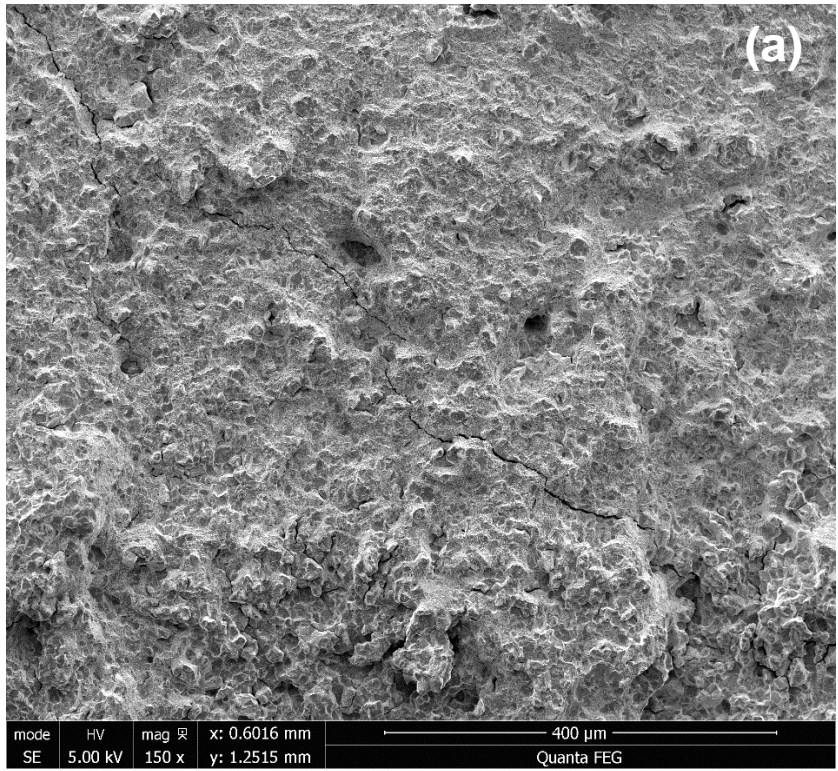

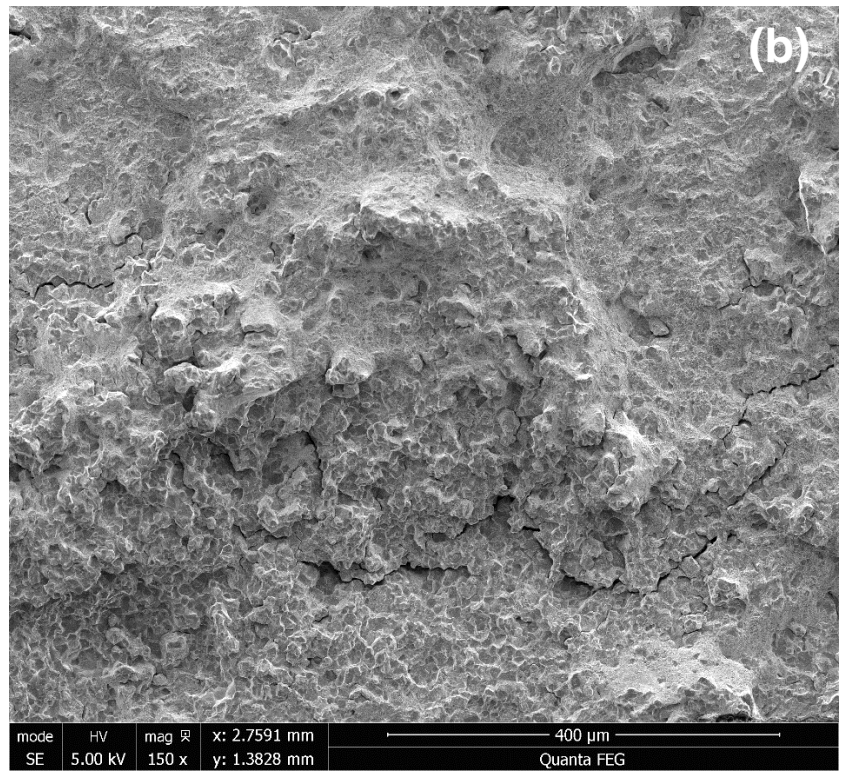

**Fig. 19.** *(a)* and *(b)* shows the fracture surfaces from the near notch tip region in 4340 at 150x.

## 3. 2. Fractography

A detailed analysis has been carried along with fracture surface mapping to develop ideas on the fracture behavior of the steel grades. The critical regions such as the 'near notch region' (where the crack initiates) and the 'center' of the fracture surface typically obtained from a 4-point bend test are shown in Fig. 20.

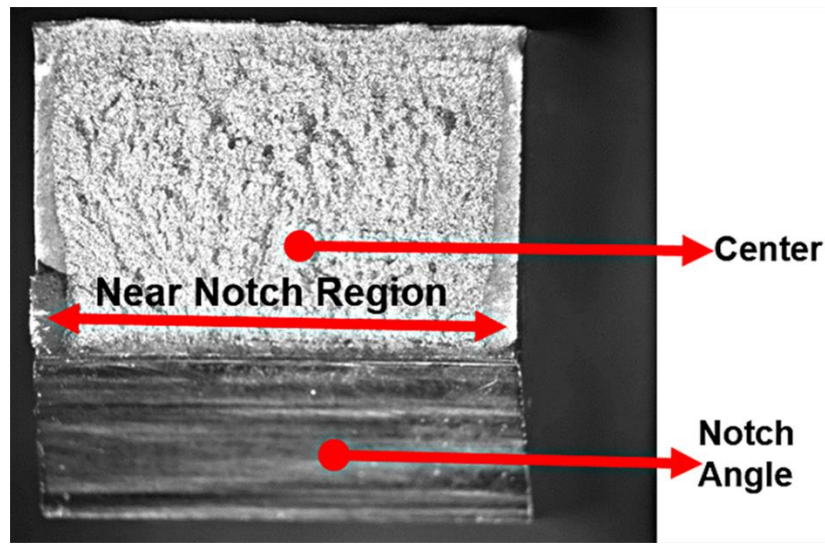

**Fig. 20.** Shows a typical fracture surface from a 4-point bend test with the critical regions (adapted from the author's prior publication in [13]).

Figure 21 shows the fracture surfaces of 4340L and 4340 obtained from the near notch region after a fast fracture test without H, at low and high magnifications respectively. It can be observed that Fig. 21 represents the usual fracture in rapid ductile failure mode, characterized by ductile-dimple morphology for all the three steels. Figures 18 and 19 (in *section 3.1.*) show the fracture surfaces of 4340L and 4340 obtained from the near notch region tested with precharged H, at a low magnification of 150x in a wider field of view. These images clearly indicate H induced fracture with conventional intergranular features at the near notch region for both 4340L and 4340, implying significant diminution of plasticity. The distribution of intergranularity reduces on moving away from the near notch or crack initiation region. The brittle to ductile transition can be noticeably observed in the images. High magnification images for 4340L are also in shown in Fig. 22.

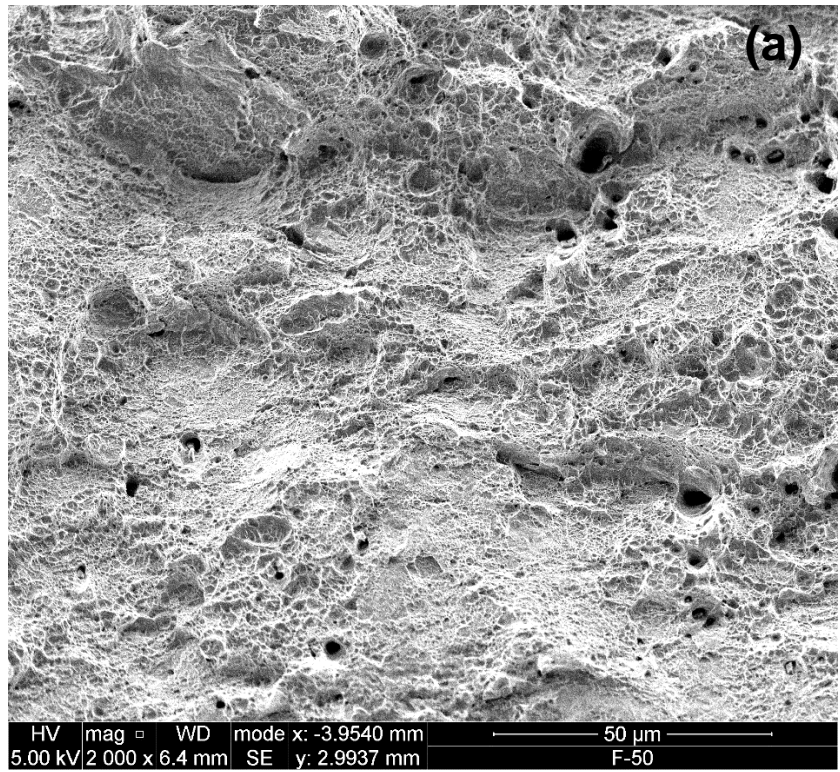

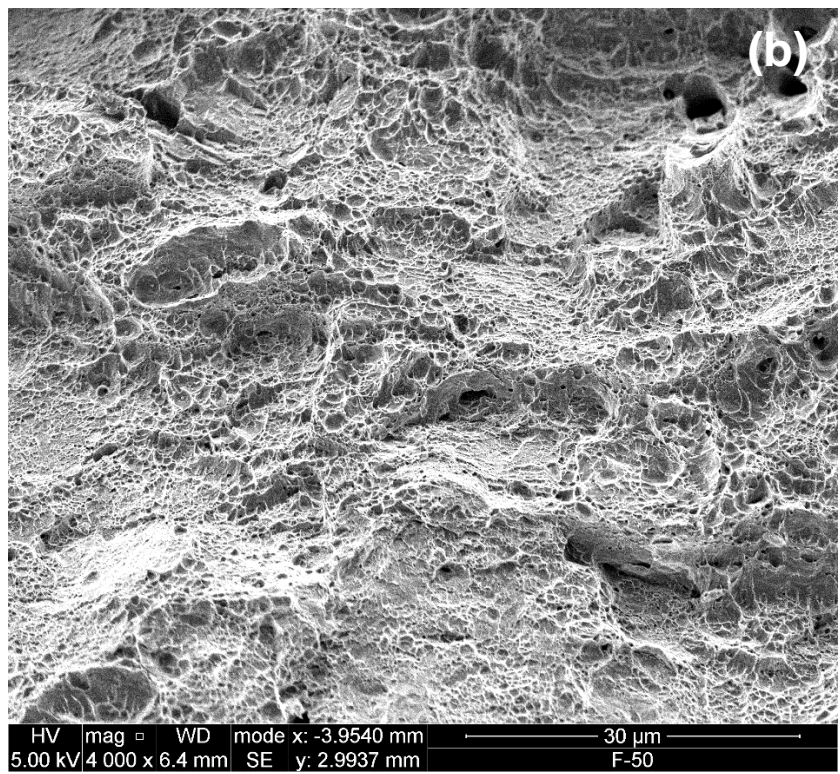

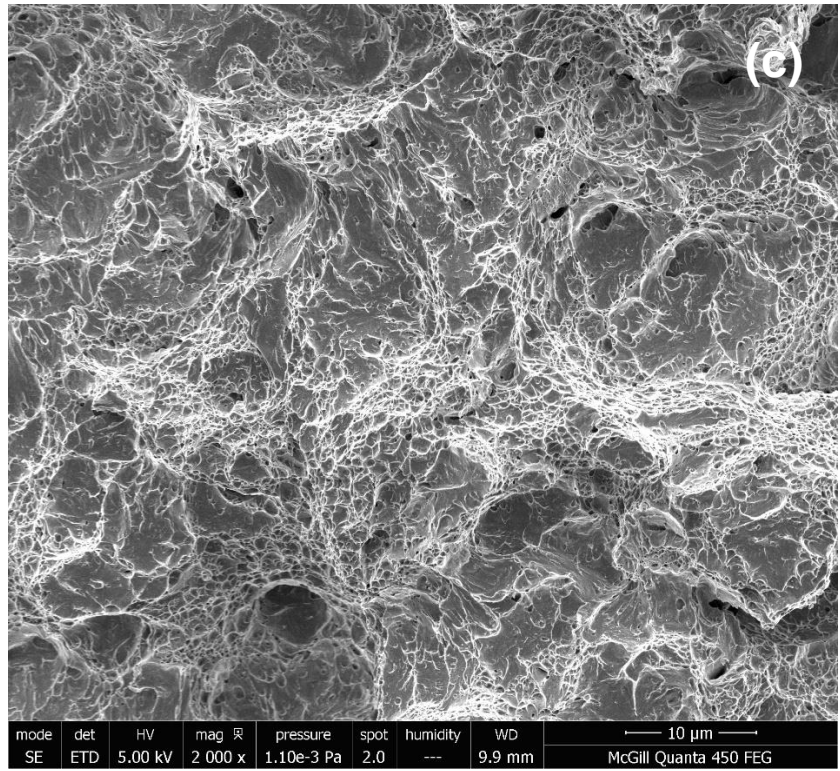

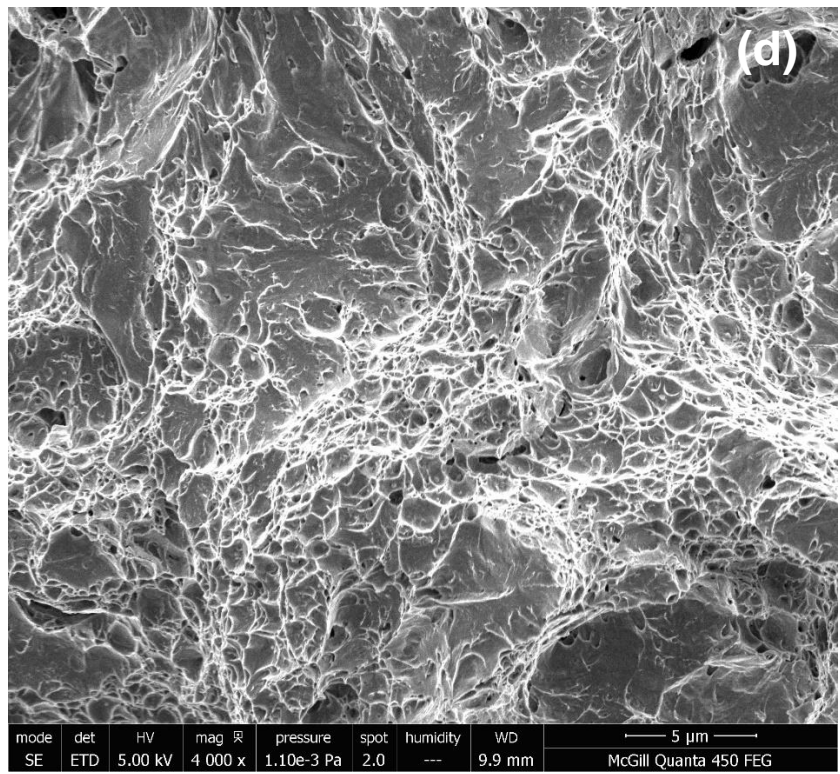

**Fig. 21.** *(a)* and *(b)* (adapted from the author's prior publication in [13]) show the fracture surfaces for 4340L, while *(c)* and *(d)* show the fracture surfaces for 4340, at 2000 x and 4000 x respectively; obtained by testing the material in air.

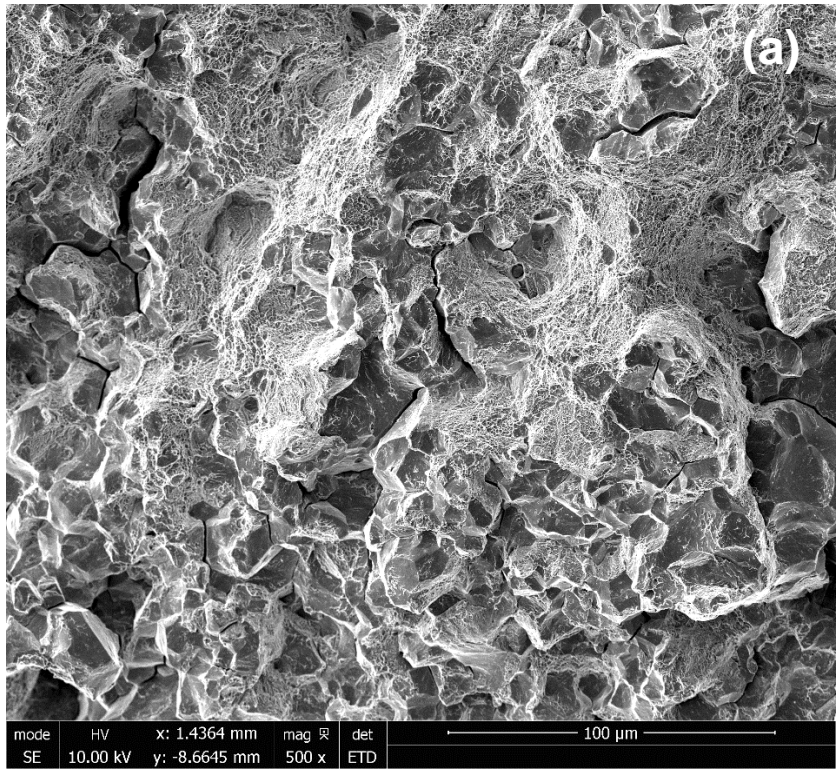

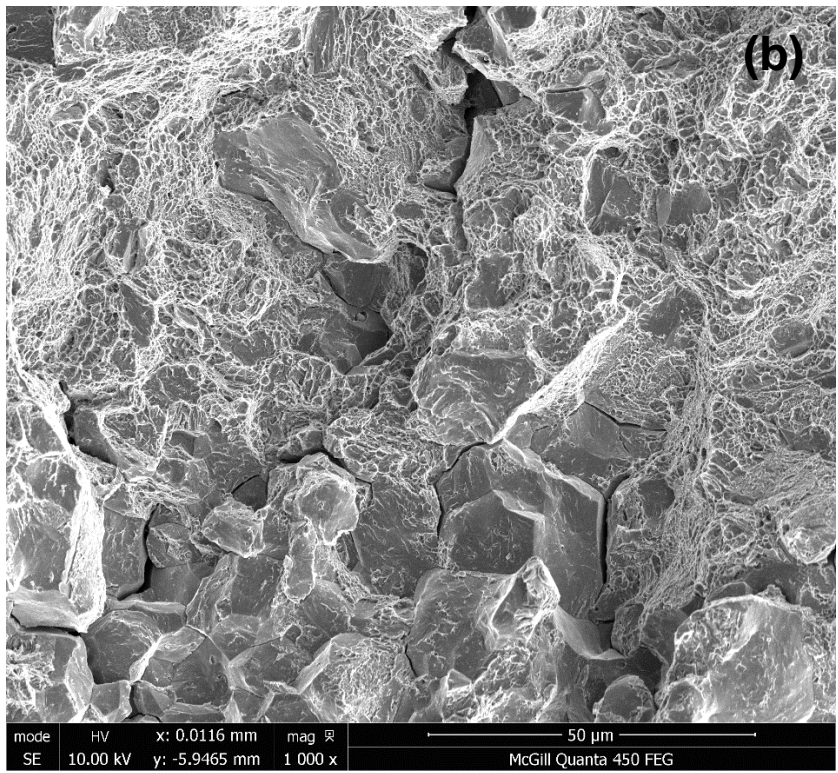

**Fig. 22.** *(a)* and *(b)* show the coexistence of ductile brittle features in the fracture surfaces of 4340L at a relatively high magnification of 500x and 1000x respectively, generated by testing the material with precharged H.

The intergranular fracture (from Fig. 18 and 19) is accompanied by a large crack running in parallel to the notch. It is also worth mentioning that the large crack formations are primarily associated with intergranularity at the crack initiation region. Small secondary cracks associated with a primary crack are observed in those areas. The secondary cracks run along the grain-boundary-like features, as visible from the high magnification images in Fig. 23, showing the severity of intergranular HE failure. On following the large cracks in the near notch region, it can be observed that they align themselves perpendicularly with respect to the notch area (i.e. parallel to the shear lips) running through the fracture surface to the other end as shown in Figs. 24 and 25 for 4340L and 4340 respectively. Interestingly, the cracks are primarily accompanied by ductile features as observed from the fracture surface. However, intergranular morphology exists intermittently (encircled in Figs. 24 and 25) and likely exist beneath the surface, which would require further investigation. Nevertheless, the mechanism of this particular type of crack formation may be explained based on the micro-mechanism proposed in *section 3.1*, where cracks can initiate on multiple planes resulting in the generation of the in-pane as well as out of plane cracks. Once the crack initiates from the notch tip (as characterized by intergranular morphology), it achieves a very high velocity partially owing to the dynamic nature of loading in this case. Additionally, hydrogen induced crack formations in high strength steels are also associated with high velocity as documented in [67,81]. Therefore, the cracks accompanied by such velocity, can penetrate through the matrix even without any presence of H, which has been also reported in [81]. Even though, these cracks could be arrested by the plasticity associated with the deformation of the matrix. But, in case of 4340L and 4340, it appears the matrix did not have enough degree of plasticity to arrest the crack, resulting in the rapid failure of those materials.

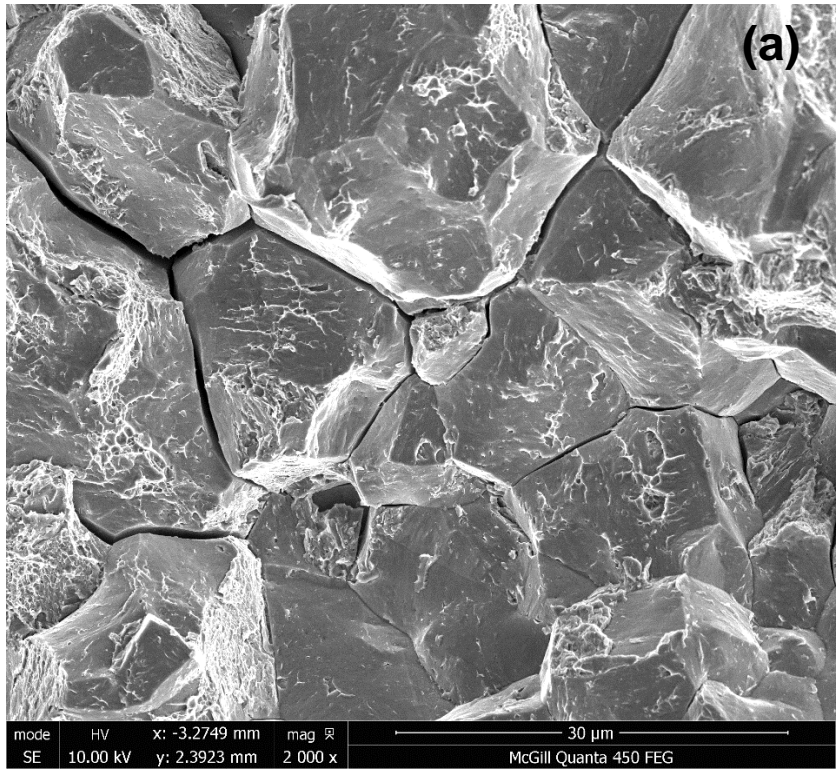

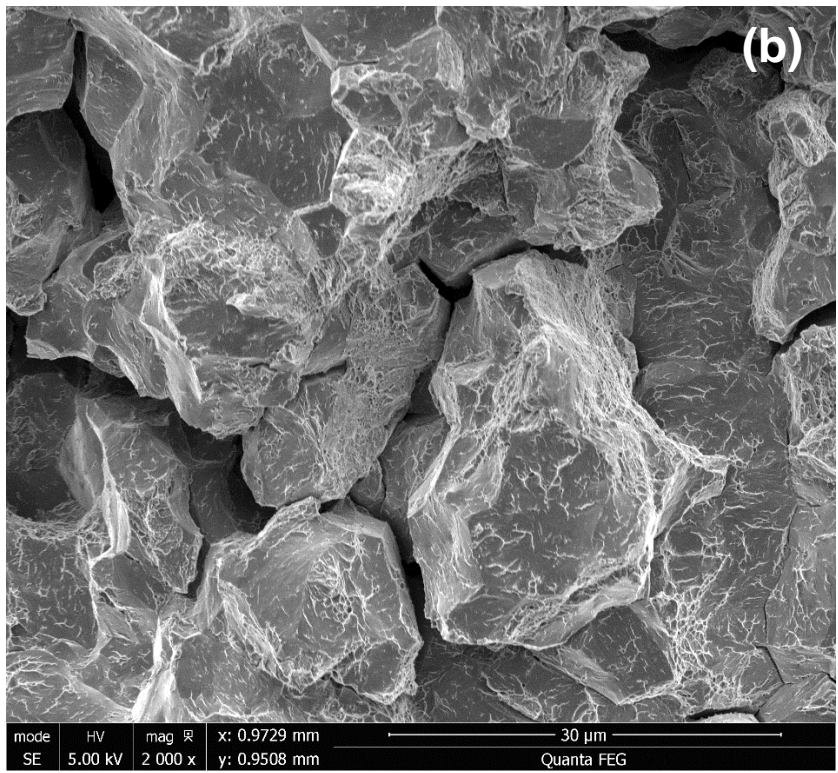

**Fig. 23.** Shows the secondary crack in (a) 4340L and (b) 4340 at a magnification of 2000x, generated while testing the material with precharged H using the fast fracture test.

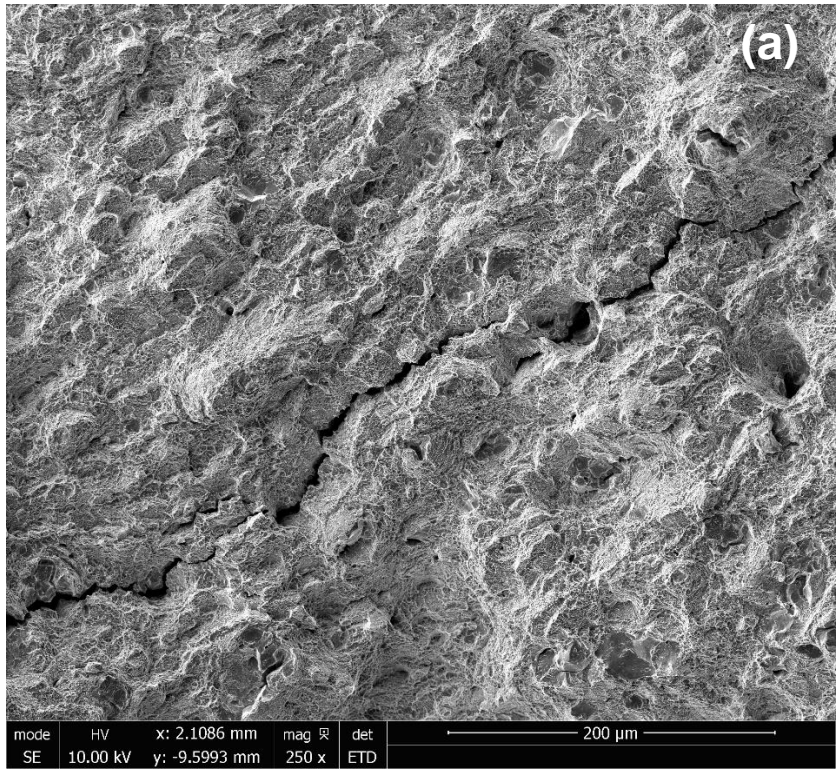

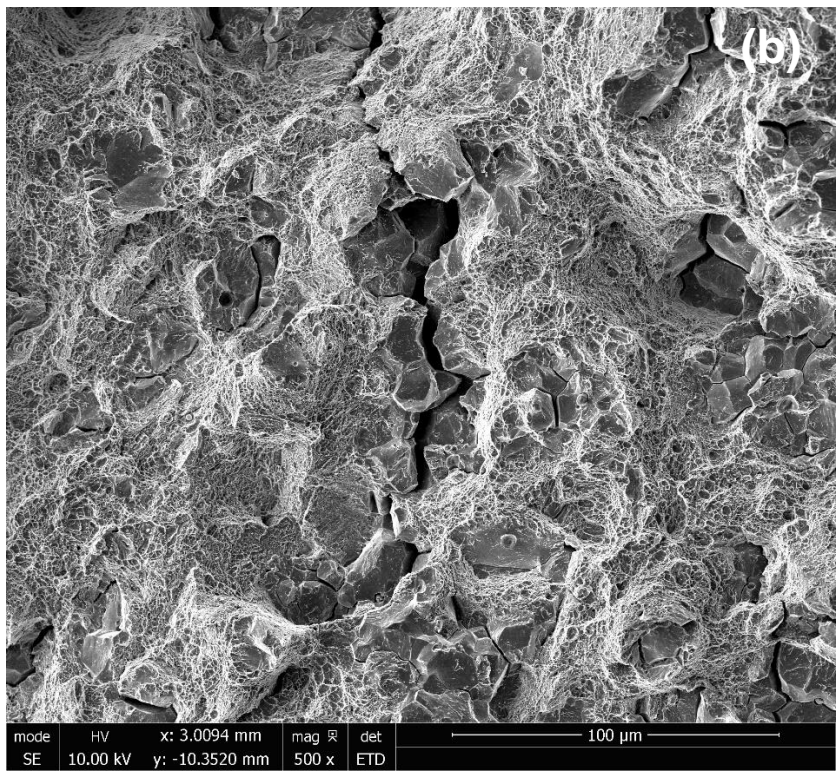

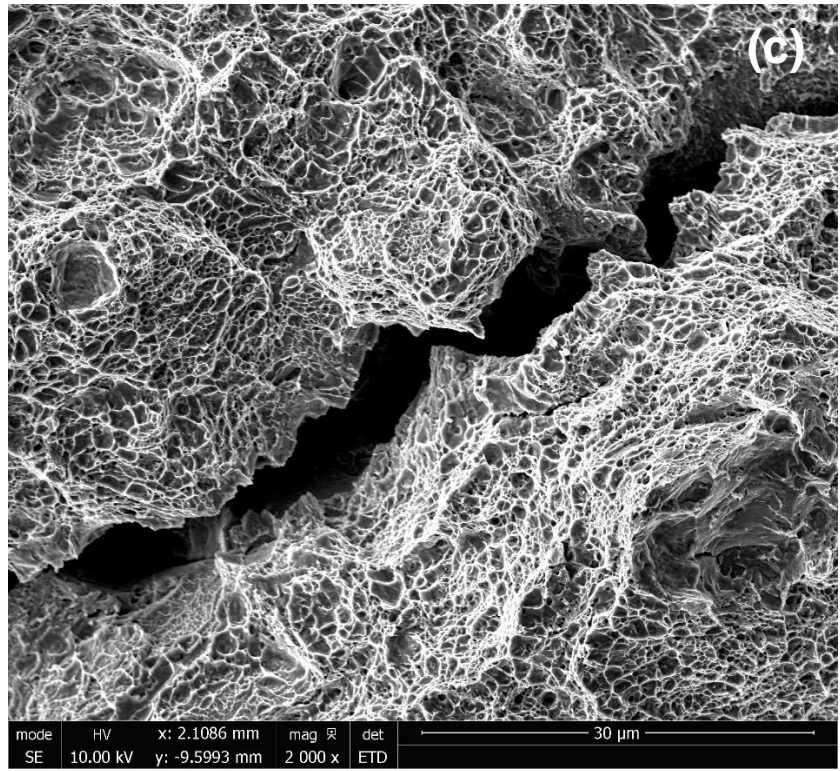

**Fig. 24.** (a), (b) and (c) show the large cracks parallel to the shear lips in 4340L at 250x, 500x and 2000x magnifications respectively, generated while testing the material with precharged H.

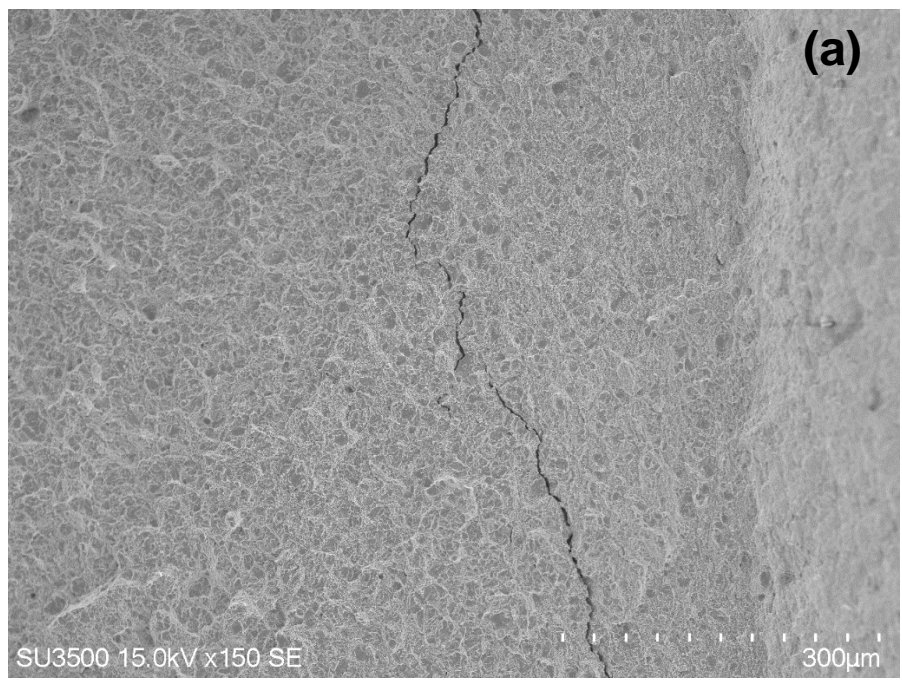

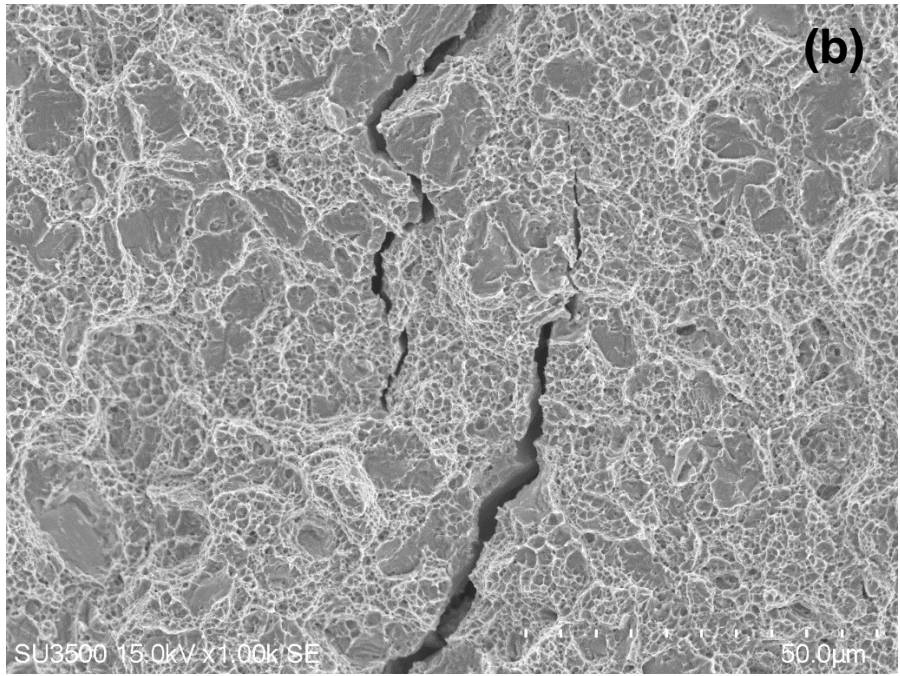

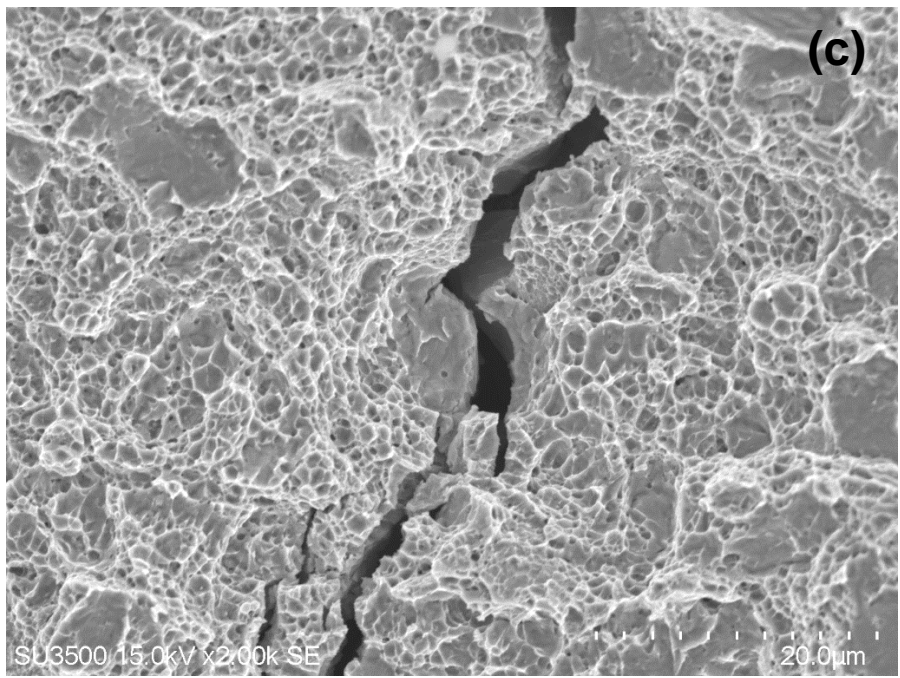

**Fig. 25.** *(a)*, *(b)* and *(c)* show the large cracks parallel to the shear lips at 150x, 1000x and 2000x magnifications respectively, generated while testing the material with precharged H.

In contrast to 4340L and 4340, no such crack formation was observed for 4140. Moreover, the fracture morphology from the near notch region in 4140 is also significantly different as compared to 4340L and 4340. Figure 26 shows the fracture surfaces from the near notch region in case of 4140, which is markedly different from 4340 and 4340L. Small pockets of intergranular features surrounded by ductile morphology can be noticed without any large scale distribution of intergranularity. This observation suggests the difference in the nature of crack formation among the materials with lower severity of HE failure in 4140, thus corroborating the observation from the novel fast fracture test methodology. Therefore, owing to the nature of crack initiation and adequate plasticity generated within the matrix (characterized by the ductile morphology), no such cataclysmic crack formations were observed in this material. However, a comprehensive discussion based on micromechanics and advanced microstructural characterization, explaining the significant difference in HE fracture behaviors between 4140 and 4340L will be presented in a successive study.

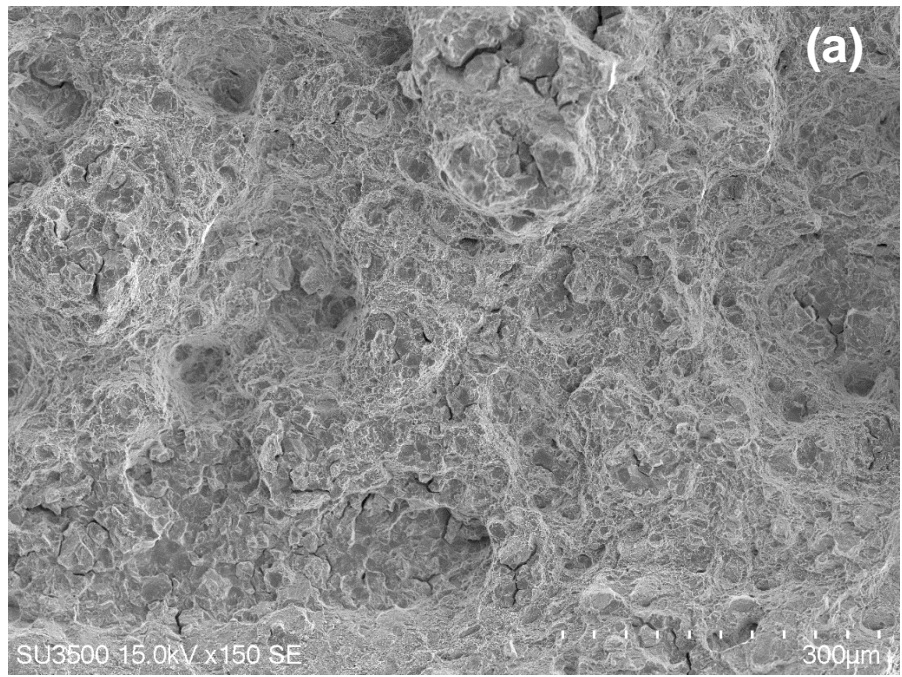

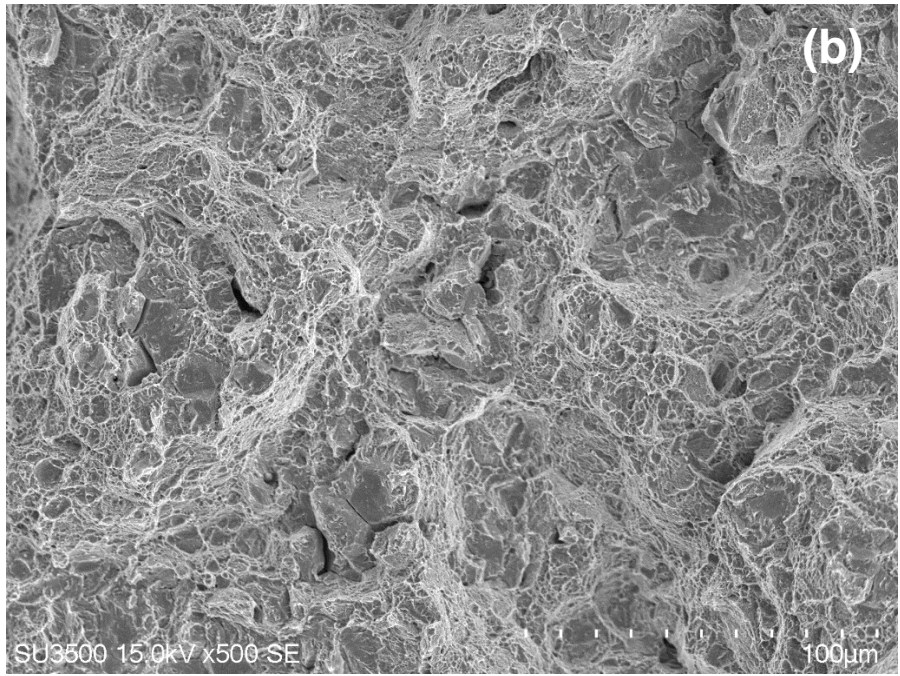

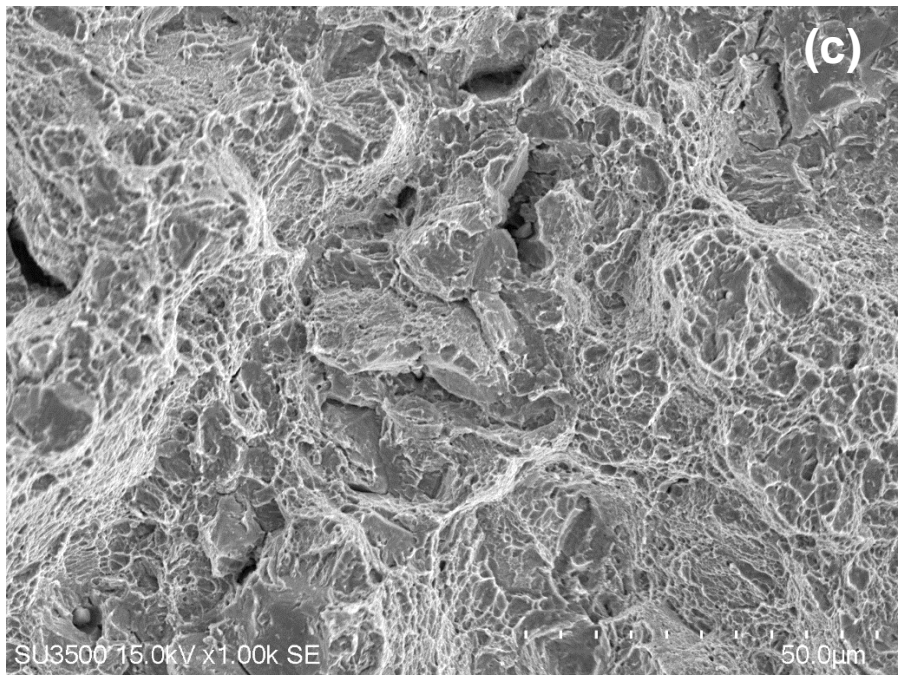

**Fig. 26.** *(a)*, *(b)* and *(c)* show the fracture surfaces from the near notch tip region in 4140, generated while testing the material with precharged H using the novel fast fracture test.

## 4. Summary

In this study, three different Q & T steels with one pair having similar steel chemistry and different strength levels, and another pair having different steel chemistries with the same strength level, were selected. A novel test, based on four point bending, with significantly high loading rate as compared to conventional HE test methods, was employed to study the HE failure mechanisms in these materials, along with FEA analyses, analytical modeling and fractography. Some of the important conclusions from the study are summarized.

1. As one of the primary results, the study offers a fundamental depiction of the role of strength on HE failure in Q & T steels, through a newly proposed HE test methodology. Two Q & T steels, 4340L and 4340, with significantly different strengths, exhibited similar fracture loads, despite exposure to the same H concentration. A mechanistic description attributes this behavior to their incapacity of deforming plastically through dislocation emissions prior to the occurrence of Griffith cleavage fracture along crystallographic planes, in the presence of H.

2. On the other hand, alloys 4140 and 4340L, quenched and tempered to the same strength level, exhibited different resistance to HE fracture. Alloy 4140 exhibited significantly lower HE susceptibility. Stress-coupled H diffusion finite element analyses showed that 4140 does exhibit sufficient degree of plastic deformation compared to 4340 and 4340L, even when exposed to same H concentration. It was further shown that dislocations generated by plastic deformation of 4140 can trap significantly more hydrogen than 4340L. These observations highlight the importance of plastic deformation toward resisting HE. However, a micromechanical description resolving further understanding is the aim of a subsequent study.

3. Finally, fractographic observations corroborate the results and discussion from the novel HE susceptibility test. The presence of predominantly intergranular fracture surface in the near notch tip area of 4340 and 4340L is consistent with the degree of embrittlement of these materials. Conversely, markedly less intergranularity was observed for 4140, reflecting a lower degree of embrittlement. Moreover, large out of plane cracks were also observed for both 4340L and 4340. This observation is also reflective of the severity of HE fracture, showing that hydrogen can increase the propensity for simultaneous decohesion along multiple planes. Similar out of plane crack formation were not observed for alloy 4140, indicating that it may be a more suitable choice for critical engineering applications in a hydrogen environment.

**Acknowledgements**

The authors would like to acknowledge the financial support from the Government of Canada through Natural Sciences and Engineering Research Council strategic grant (NSERC STPGP 521860-18), to carry out this research work. TD acknowledges the financial support from Dr. Y Lin-Alexander fellowship under the McGill Engineering Doctoral award (MEDA) program. TD also acknowledges Dr. Sriraman Rajagopalan for helping him in setting up the hydrogen charging workstation. TD would also like to thanks Dr. Ammar Alsheghri for helpful discussions on computational techniques.

**Appendix A**

*2D finite element diffusion analysis*

Finite element simulation was performed with a constant H concentration boundary condition in a single-edge notched SE(B) specimen (Fig. 1) to study the evolution of lattice H distribution along the notch length over a period of 1 hour as shown in Fig. A.1. The boundary H concentration was 36.305 mol/m$^3$ (~4.6 wt. ppm) calculated from Eq. (11).

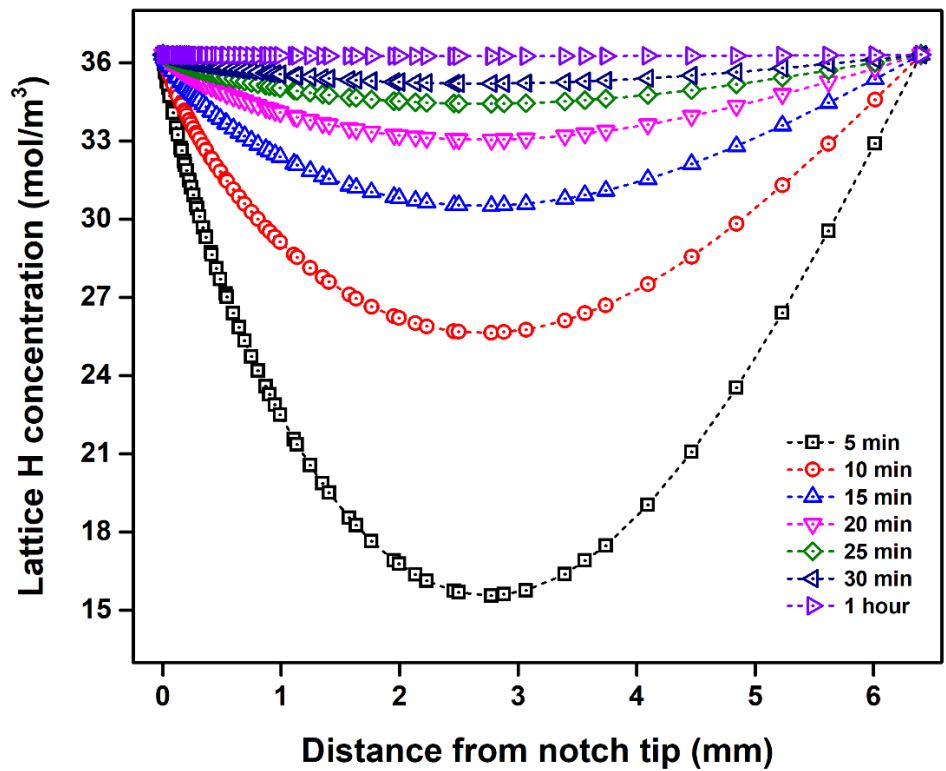

**Fig. A.1.** Shows the evolution of H concentration along the notch length of the single-edge notched SE(B) specimen.

Based on the FEA calculations, it can be seen from Fig. A.1. that an equilibrium is reached with the boundary (or surface) concentration after 30 mins of H charging over a total time of 1 hour charging. Thus, a uniform hydrogen distribution in the near notch area is highly feasible. The calculation further justifies the assumption of an initial condition with uniform hydrogen distribution for the stress coupled FEA simulations.

**Data Availability Statement**

The data generated during the current study can be made available only upon reasonable request.